\pdfoutput=1
\documentclass[10pt]{article}
\usepackage[accepted]{tmlr}

\usepackage{amsmath,amsfonts,bm}

\def\eqref#1{equation~\ref{#1}}

\def\1{\bm{1}}

\DeclareMathAlphabet{\mathsfit}{\encodingdefault}{\sfdefault}{m}{sl}
\SetMathAlphabet{\mathsfit}{bold}{\encodingdefault}{\sfdefault}{bx}{n}

\usepackage{graphicx}
\graphicspath{ {figures/} }

\usepackage{hyperref}
\usepackage{url}

\usepackage{amsmath,amssymb,amsfonts,amsthm}
\usepackage[algo2e,ruled,lined,boxed,commentsnumbered, noend, linesnumbered, procnumbered]{algorithm2e}
\usepackage{cleveref}
\usepackage{siunitx}
\usepackage{booktabs} 
\usepackage{multirow}
\usepackage{tikz}
\usepackage{diagbox}
\usepackage{makecell}
\usepackage[inline]{enumitem}

\usepackage[abbreviations]{foreign}
\usepackage{colortbl}
\usepackage{float}
\restylefloat{table}
\usepackage{threeparttable}
\usepackage{tabularx}

\usepackage{xspace}

\newcommand{\cifar}{CIFAR-10\xspace}
\newcommand{\tinydataset}{Tiny\xspace}
\newcommand{\imagenet}{ImageNet\xspace}
\newcommand{\resnettwenty}{ResNet-20\xspace}
\newcommand{\resneteighteen}{ResNet-18\xspace}
\newcommand{\resnetthirtyfour}{ResNet-34\xspace}
\usepackage{tikz}
\usepackage{pgfplots}
\usepackage{pgfplotstable}
\ifdefined\tikzexternalrealjob\renewcommand{\sfdefault}{phv}\fi
\usepackage[eulergreek]{sansmath}
\usepackage{comment}
\newcolumntype{L}[1]{>{\raggedright\let\newline\\\arraybackslash\hspace{0pt}}m{#1}}
\newcolumntype{C}[1]{>{\centering\let\newline\\\arraybackslash\hspace{0pt}}m{#1}}
\newcolumntype{R}[1]{>{\raggedleft\let\newline\\\arraybackslash\hspace{0pt}}m{#1}}

\pgfplotsset{compat=newest}
\usepgfplotslibrary{external,units,colorbrewer,groupplots,fillbetween}
\tikzsetexternalprefix{figures/}
\tikzset{external/mode=list and make}
\usetikzlibrary{patterns}
\usetikzlibrary{shapes.geometric, arrows, positioning}

\newcommand{\newgroupwidth}[2]%
{\expandafter\xdef\csname groupwidth#1\endcsname{#2}}

\newcounter{groupwidth}
\newsavebox{\groupwidthbox}
\makeatletter
{\edef\groupnumber{#1}%
	\stepcounter{groupwidth}%
	\@ifundefined{groupwidth\thegroupwidth}{\pgfmathsetlengthmacro{\mywidth}{\linewidth/\groupnumber}}%
	{\expandafter\let\expandafter\mywidth\csname groupwidth\thegroupwidth\endcsname}%
	\begin{lrbox}{\groupwidthbox}%
		\tikzset{/pgfplots/width={\mywidth}}%
		\ignorespaces}%
	{\end{lrbox}%
	\usebox\groupwidthbox
	\pgfmathsetlengthmacro{\mywidth}{\mywidth + (\linewidth - \wd\groupwidthbox)/\groupnumber}
	\immediate\write\@auxout{\string\newgroupwidth{\thegroupwidth}{\mywidth}}}
\makeatother
\usepackage{acronym}
\acrodef{ML}{machine learning}
\acrodef{FHE}{fully homomorphic encryption}
\acrodef{CNN}{convolutional neural network}
\acrodef{LWE}{learning with errors}
\acrodef{RLWE}{ring learning with errors}
\acrodef{DP}{differential privacy}
\acrodef{TFHE}{fully homomorphic encryption over the torus}
\acrodef{MLaaS}{ML-as-a-Service}
\acrodef{E2E}{end-to-end}
\acrodef{QAT}{quantization-aware training}
\newcommand{\sys}{\textsc{Safhire}\xspace}
\newcommand{\orion}{\textsc{Orion}\xspace}
\newcommand{\relu}{\textsc{ReLU}\xspace}
\newcommand{\hyphen}{\textsc{HyPHEN}\xspace}
\newcommand{\concreteml}{\textsc{Concrete-ML}\xspace}

\newcommand{\codeurl}{\texttt{https://github.com/sacs-epfl/SAFHIRE}}
\usepackage{thm-restate}
\usepackage{bbm}
\newtheorem{theorem}{Theorem}
\newtheorem{lemma}[theorem]{Lemma}

\newtheorem{remark}[theorem]{Remark}

\usepackage{physics} 
\usepackage{algpseudocode}
\usepackage{amsmath}
\usepackage{booktabs}
\usepackage{multirow}
\usepackage{graphicx}
\allowdisplaybreaks

\title{Practical and Private Hybrid ML Inference with\\Fully Homomorphic Encryption}

\author{\name Sayan Biswas \email sayan.biswas@epfl.ch\\
    \addr EPFL
    \AND
    \name Philippe Chartier \email philippe.chartier@inria.fr\\
    \addr Inria, IRMAR, Universit\'e de Rennes
    \AND
    \name Akash Dhasade \email adhasade@andrew.cmu.edu\\
    \addr Carnegie Mellon University
    \AND
    \name Tom Jurien \email tjurien@stanford.edu\\
    \addr Stanford
    \AND
    \name David Kerriou \email david.kerriou@polytechnique.edu\\
    \addr \'Ecole Polytechnique
    \AND
    \name Anne-Marie Kermarrec \email anne-marie.kermarrec@epfl.ch\\
    \addr EPFL
    \AND
    \name Mohammed Lemou \email mohammed.lemou@univ-rennes.fr\\
    \addr CNRS, IRMAR, Universit\'e de Rennes
    \AND
    \name Franklin Tranie \email ftranie@ethz.ch\\
    \addr ETH Zurich
    \AND
    \name Martijn de Vos \email martijn.devos@epfl.ch\\
    \addr EPFL
    \AND
    \name Milos Vujasinovic \email milos.vujasinovic@epfl.ch\\
    \addr EPFL
}

\def\month{09}
\def\year{2026}
\def\openreview{\url{https://openreview.net/forum?id=AAneiBMYgS}}

\begin{document}

\maketitle

\begin{abstract}
In contemporary cloud-based services, protecting users' sensitive data and ensuring the confidentiality of the server's model are critical.
Fully homomorphic encryption (FHE) enables inference directly on encrypted inputs, but its practicality is hindered by expensive bootstrapping and inefficient approximations of non-linear activations.
We introduce \sys, a hybrid inference framework that executes linear layers under encryption on the server while offloading non-linearities to the client in plaintext.
This design eliminates bootstrapping, supports exact activations, and significantly reduces computation cost.
To safeguard model confidentiality despite client access to intermediate outputs, \sys applies randomized shuffling, which obfuscates intermediate values and makes it practically impossible for a client to reconstruct the model.
To further reduce inference latency, \sys incorporates advanced optimizations such as fast ciphertext packing and partial extraction.
Evaluations on multiple standard models and datasets show that \sys achieves up to \(7.6\times\) and \(7.4\times\)  lower inference latency than \orion and \concreteml, respectively, which are two state-of-the-art FHE baselines, with manageable communication overhead and comparable accuracy.
These results underline the practicality of our hybrid FHE inference scheme.

 \end{abstract}

\acresetall
\section{Introduction}
\label{sec:introduction}

\Ac{ML} has profoundly impacted industrial domains such as healthcare~\citep{habehh2021machine}, finance~\citep{dixon2020machine}, and the Internet of Things~\citep{hussain2020machine}.
In many practical deployments, \ac{ML} models are hosted on cloud servers and accessed by clients who submit private and privileged data for inference~\citep{zhang2019mark}.
However, this paradigm, also known as \ac{MLaaS}, raises serious privacy concerns as users must share their inference inputs, such as medical or financial data, with cloud servers who can directly use or share the data with other parties~\citep{mireshghallah2020shredder}.
As users increasingly rely on online inference services, ensuring the privacy of user data during inference is paramount~\citep{hesamifard2018privacy}.

\begin{figure}[t]
	\centering
	\includegraphics[width=.5\linewidth]{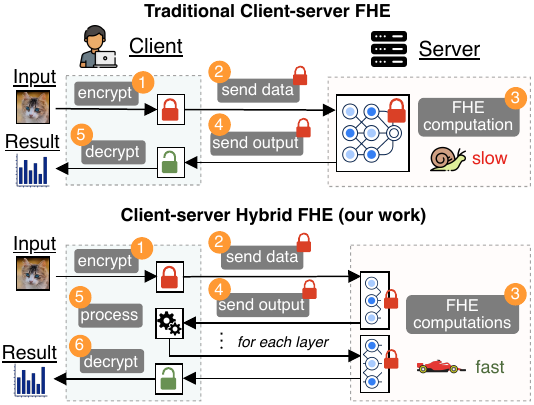}
	\caption{Existing \ac{FHE}-based \ac{ML} inference (top) performs all computation on the server in the encrypted domain; our hybrid scheme (bottom) offloads computationally expensive operations to the client to significantly improve efficiency.}
	\label{fig:fhe}
\end{figure}

\Ac{FHE} offers a compelling solution to this privacy concern by enabling computations directly on encrypted user data, see \Cref{fig:fhe} (top)~\citep{gentry2009fully,marcolla2022survey}.
\ac{FHE} provides strong privacy guarantees as the server learns nothing beyond ciphertext sizes and protocol metadata.
An \ac{ML} inference request with \ac{FHE} works as follows: a user first encrypts its inference inputs and sends the encrypted data to the server (step 1–2).
The server then performs the computations related to the neural network forward pass in the encrypted domain (step 3) and sends back the encrypted output (step 4), after which the user decrypts the output and obtains the inference result (step 5).
This way, the server is unable to infer any information from the input data since it remains encrypted during the inference request~\citep{cryptonets}.

\begin{figure}[t!]
	\centering
	\includegraphics{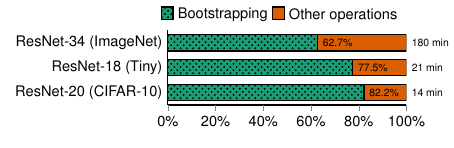}
	\caption{Latency breakdown for bootstrapping and other operations during a single sample inference request using \orion, a state-of-the-art \ac{FHE} inference framework. We evaluate three models and datasets and show the \ac{E2E} inference duration in minutes on the right.}
	\label{fig:motivation_bootstrapping_time}
\end{figure}

Existing \ac{FHE} inference schemes face two main obstacles that make \ac{E2E} encrypted inference impractical at scale~\citep{orion,nam2022accelerating}.
First, encrypted vectors can only undergo so many operations before becoming too noisy to decrypt correctly.
To reset ciphertext noise, \ac{FHE} schemes rely on an operation called \emph{bootstrapping}.
Bootstrapping enables further computations on these vectors by reducing this noise, but comes with a high computational cost~\citep{bossuat2021efficient}.
\Cref{fig:motivation_bootstrapping_time} shows the duration of bootstrapping and other operations in \orion, a state-of-the-art \ac{FHE} inference framework~\citep{orion}.
We evaluate the latency required for bootstrapping and other operations for three \ac{CNN} models, \resnettwenty, \resneteighteen and \resnetthirtyfour, and three datasets, \cifar, \tinydataset and \imagenet.
Across these configurations, bootstrapping takes between 63\% to 82\% of total inference time, highlighting the significant cost of this operation.
Second, some non-linear operations commonly included in neural networks, such as \relu, are not natively supported by \ac{FHE} and must be replaced by, for example, high-order polynomial approximations or softmax functions~\citep{daubechies2022nonlinear,zhang2024efficient,ishiyama2020highly}.
Unfortunately, these approximations require many more computations than their native counterparts and quickly accumulate ciphertext noise, necessitating additional bootstrapping steps and thus increasing latency.

This work introduces \sys, a novel and practical hybrid \ac{FHE} scheme for \ac{ML} inference that overcomes the above limitations.
\sys leverages a simple yet powerful idea: we keep the computation of linear, parameterized layers (\eg, convolutions and fully connected layers) on the server under encryption and push non-linear operations (\eg, ReLU activations) to the client which evaluates them in plaintext.
Operationally, \sys proceeds in rounds that correspond one-for-one to linear blocks bounded by non-linearities.\footnote{While we primarily evaluate on CNNs (in particular ResNet models), our approach applies to any architecture that alternates linear operations with non-linear ones (\eg, blocks that are linear up to an activation).}
We show the \ac{E2E} workflow of \sys in \Cref{fig:fhe} (bottom).
In each round, the client encrypts and sends the current layer input (steps 1–2), the server applies the encrypted linear transform (step 3) and returns the resulting ciphertexts (step 4). Upon receipt, the client decrypts and applies the exact non-linear operation (\eg, \relu), then re-encrypts the outputs for the next round (step 5).
This process repeats until the forward pass completes after which the client obtains the final result (step 6).
The intermediate decryption and re-encryption at the client helps avoid bootstrapping since we decrypt before the noise becomes unmanageable.
Moreover, because all non-linearities are executed in plaintext on the client, \sys avoids any server side non-linear approximations, preventing additional bootstrapping and amplified inference latency. 

However, revealing intermediate outputs to the client with \sys poses the risk of model reconstruction wherein the client may try to reverse engineer the model weights. 
This undermines \textit{model confidentiality} as the server model often represents proprietary intellectual property and investment due to high training and deployment costs.
To prevent this, \sys randomly shuffles the outputs of the linear operations on the server before sending them to the client.
The shuffling function and the associated unshuffling function is derived by the server from a per-session secret seed and is unknown to the client.
We provide formal differential privacy guarantees resulting from the shuffling operation in combination with the default noise introduced by server side operations.
Moreover, \Cref{sec:model_privacy} shows that in \sys the information an adversarial client can recover about the server's model is limited to a permuted weight matrix per layer, or its histogram when the server both shuffles a layer's outputs and unshuffles its inputs.
At the same time, the correctness of the client-side decryption remains unaffected.

Under the hood, \sys leverages the \ac{TFHE}-based \ac{RLWE} 
cryptographic scheme which is widely used for practical homomorphic computation~\citep{tfhe2018,cryptoeprint:2025/488}.
A key advantage of the \ac{TFHE} scheme is that it allows us to apply fine-grained advanced optimizations, such as fast ciphertext packing and
partial trace extraction, resulting in substantial performance improvements.
Our optimizations reduce both computation and communication costs, yielding practical \ac{E2E} latency under standard security parameters.
We implement \sys and publish its code as a library on GitHub.\footnote{The code of SAFHIRE is available at \codeurl{}}
Our implementation supports inference using multi-core CPUs and GPUs.

We evaluate the efficiency of our scheme using widely adopted \acp{CNN} architectures and different datasets.
Under realistic network conditions, \sys using a single thread reduces \ac{E2E} latency of inference requests by up to 7.6$\times$ compared to \orion, a state-of-the-art single-threaded baseline.
When running \sys with 32 threads, we obtain up to 7.4$\times$ reduction in \ac{E2E} compute inference latency compared to \concreteml, a multi-threaded baseline that also offloads activations to the client.
This comes at a manageable ciphertext communication cost of at most \SI{499}{\mega\byte} for \resneteighteen and \SI{170}{\mega\byte} for \resnettwenty.
Moreover, we show that through the use of multithreading \sys reduces server-side execution time up to 54$\times$ compared to \orion.
Using GPU acceleration, our inference duration on \cifar using model \resnettwenty can be as low as 25.7 seconds even with a modest 1.25\si{\mega\byte\per\second} internet connection.
In summary, \sys offers a practical step toward efficient \ac{FHE} inference with high model accuracy.

\textbf{Contributions.} Our key contributions are as follows:
\begin{itemize}
  \item We introduce \sys, a hybrid and efficient \ac{FHE} inference framework that executes linear, parameterized layers on the server under encryption while offloading non-linearities to the client in plaintext, eliminating bootstrapping while supporting non-linearities without approximations (\Cref{sec:protocol}). 
  \item We show that, with randomized shuffling, \sys helps preserve server-side model confidentiality, in addition to enabling clients to perform model inference in a private and secure manner. We derive the amplified differential privacy guarantees, jointly emerging from shuffling with inherent ciphertext noise (\Cref{sec:model_privacy}).
  \item We implement \sys using the \ac{TFHE}-based \ac{RLWE} scheme 
  and develop efficiency optimizations, including high-throughput ciphertext packing and partial trace extraction, greatly reducing computational cost.
  \item We evaluate \sys on standard \ac{CNN} model architectures and datasets, demonstrating up to 7.6$\times$ lower \ac{E2E} latency and up to 63$\times$ less server compute than \orion depending on the dataset and configuration per inference with comparable accuracy (\Cref{sec:evaluation}).
\end{itemize}

\section{Background and Preliminaries}
\label{sec:preliminaries}

\sys builds on a small set of primitives from \ac{TFHE}. This section introduces them in the order in which they build on one another. For each primitive, we first explain why \sys needs it, then define it, and finally point to the steps of the workflow in \Cref{fig:workflow} and to the parts of \Cref{sec:protocol} where it is used. Appendix~\ref{app:fhe-intuition} complements this section with an example of one round of working of \sys consolidating all these primitives. \cite{cryptoeprint:2025/488} provides a more formal and detailed explanation of these concepts. A summary of the key notations used in this work is provided in Appendix~\ref{sec:app:table_of_notation}. 
 
\paragraph{How the primitives fit together.}
In each round of \sys (\Cref{fig:workflow}), the client encrypts many layer inputs into a few \ac{RLWE} ciphertexts and sends them to the server (steps~2-5). The server \emph{unpacks} these ciphertexts, so that each input value is carried by its own ciphertext (step~6), evaluates the linear layer on them with its plaintext weights (step~8), and \emph{packs} the outputs back into a few ciphertexts (step~10), which the client decrypts (step~12). This requires an encryption scheme that holds many values per ciphertext and supports linear operations (\Cref{sec:enc_dec}), together with a way to move values out of, and back into, individual positions of a ciphertext. The latter is provided by the trace operator (\Cref{sec:fast_trace}), whose homomorphic evaluation relies on automorphisms and key switching (Section~\ref{sec:key_switch}), and on which packing is built (Section~\ref{sec:packing}). Table~\ref{tab:primitives} summarizes the role of each primitive.

\begin{table}[t]
\centering
\small
\renewcommand{\arraystretch}{1.15}
\begin{tabular}{@{}p{0.17\textwidth}p{0.42\textwidth}p{0.35\textwidth}@{}}
\hline
\textbf{Primitive} & \textbf{Why \sys needs it} & \textbf{Where it is used in \sys's workflow (\Cref{fig:workflow})} \tabularnewline
\hline
\raggedright \Ac{LWE} and \Ac{RLWE} encryption (\Cref{sec:enc_dec})
& \raggedright Hides the client's data from the server. An RLWE ciphertext carries up to $N$ values, which keeps communication low. Since the client decrypts after every layer, the noise is reset without bootstrapping.
& \raggedright Key generation; encryption, steps~2-4; decryption, step~12. \tabularnewline
\raggedright Homomorphic linear operations (\Cref{sec:enc_dec})
& \raggedright Let the server evaluate linear layers on ciphertexts, provided that the weights are integers, hence the quantized model.
& \raggedright Partial extraction, step~6, and linear layer, step~8. \tabularnewline
\raggedright Automorphisms and key switching (\Cref{sec:key_switch})
& \raggedright Needed to evaluate traces homomorphically: an automorphism changes the key under which a ciphertext decrypts, and key switching restores it.
& \raggedright Keys generated at setup; fast packing, step~10; avoided in step~6 thanks to the powers computed by the client in step~3. \tabularnewline
\raggedright Trace (\Cref{sec:fast_trace})
& \raggedright Isolates one coefficient of an encrypted polynomial. Computed level by level along a tower of subfields, which reduces its cost.
& \raggedright Power computation, step~3; partial trace extraction, step~6; fast packing, step~10. \tabularnewline
\raggedright Packing (\Cref{sec:packing})
& \raggedright Gathers the outputs of a layer, each carried by its own ciphertext, into a few RLWE ciphertexts before they are returned.
& \raggedright Fast packing, step~10, and transfer, step~11. \tabularnewline
\hline
\end{tabular}
\caption{FHE primitives used by \sys, why they are needed, and where they appear in the workflow of \Cref{fig:workflow}.}
\label{tab:primitives}
\end{table}

\subsection{Encryption and Decryption with LWE and RLWE}\label{sec:enc_dec}

\emph{\Ac{LWE}} is a cryptographic problem widely used in post-quantum cryptography due to its resilience against quantum attacks~\citep{regev2009lattices}. The \ac{LWE} problem essentially focuses on solving systems of noisy linear equations. Let us consider a discrete torus
$T_p=\frac{1}{p}\mathbb{Z}_p$ and $T_q=\frac{1}{q}\mathbb{Z}_q$ where $p$ is an integer representing the size of the message space and $q$ is the size of the ciphertext space. In this work, values for $p$ range from $2^8$ to $2^{16}$ while the value of $q$ is taken to be $2^{53}$ such that $T_q$ is representable by double floating point precision.
Besides, let $\mathbb{S}$ be a finite subset of $\mathbb{Z}$. Given a message $\mu \in T_p$ and a secret key $s\in\mathbb{S}^n$, we sample a $n$-dimensional mask $a$ uniformly at random from $T_q^{n}$ and a noise $ e \sim \mathcal{N}(0,\Delta^2)$ where $\mathcal{N}(0,\Delta^2)$ denotes Gaussian distribution with mean $0$ and standard deviation $\Delta$.
An \ac{LWE}-encryption of $\mu$ using the secret key $s$ is given by $\operatorname{LWE}_{s}(\mu) = (a, b)$, where $b = \sum_{i=1}^{n} s_i a_i+ \mu + e\mod 1$. 
The corresponding \ac{LWE}-decryption of $(a,b)$ with a secret key $s$ is given by
$\pi_p(b - \sum_{i=1}^{n} s_i a_i)$ where $\pi_p$ is a projection on the discrete torus $T_p$ defined as $\pi_p(\xi)=\frac{\lfloor p\xi\rceil}{p}$ for any $\xi\in\mathbb{R}$. The decryption is correct as long as $\left|e \right| < 1/2p$.

The above encryption-decryption scheme is canonically extended to polynomials in the cyclotomic quotient ring $T_q[X]/\Phi_M[X]$ where $\Phi_M$ is a prime-power cyclotomic polynomial (\ie $M=t^\alpha$ for a prime $t$ and a non-negative integer $\alpha$). This gives rise to \emph{\acf{RLWE}}. Given a polynomial $\mu[X]\in T_p[X]/\Phi_M[X]$ and a secret polynomial key $s(X)\in \mathbb{S}[X]$ of degree $N=\phi(M)$ (here $\phi$ denotes the Euler's totient function), we form $a^*(X)=\sum_{i=0}^{N-1} a_i^*\Omega_i^*(X)$, where each $a^*_0,\ldots a^*_{N-1}$ is sampled uniformly from $T_q$, and $e^*(X)=\sum_{i=0}^{N-1} e_i^*\Omega_i^*(X)$ where $e^*_i\sim \mathcal{N}(0,\Delta^2)$ for all $i=1,\ldots,N-1$. Here, the family $(\Omega^*_i)_{0\leq i \leq N-1}$ is meant to be the dual basis of $(X^i)_{0\leq i \leq N-1}$ with respect to an appropriate scalar-product.
An \ac{RLWE}-encryption of $\mu(X)$ using the secret key polynomial $s(X)$ is given by $\operatorname{RLWE}_{s(X)}(\mu(X))=(a(X),b(X))$ where 
$a(X) = (\Omega^*_0)^{-1}(X) a^*(X)$ and 
$b(X) = s(X) a(X) + \mu(X) + (\Omega^*_0)^{-1}(X) e^*(X)$.
Similar to \ac{LWE}, the decryption is performed by applying $\pi_p$ coefficient-wise to $b(X)-s(X)\cdot a(X)$ and it is correct if $\norm{(\Omega^*_0)^{-1}(X)e^*(X)}_{\infty} < 1/2p $.

\subsection{Key-switching}\label{sec:key_switch}
\emph{Key-switching} is a technique that changes an \ac{RLWE} ciphertext
of a given message encrypted under one key to another \ac{RLWE} ciphertext of the same message encrypted under another key.
Given an \ac{RLWE} encryption of $\mu(X)$ with a secret key $s(X)$, and a key-switching key $\operatorname{KSK}_{s(X) \rightarrow s'(X)}$, the key-switching operation outputs $\operatorname{RLWE}_{s'(X)}(\mu(X))$,
an \ac{RLWE} encryption of $\mu(X)$ with secret key $s'(X)$. 
In \ac{RLWE} schemes, ring automorphisms are applied to ciphertexts, which effectively change the secret key from $s(X)$ to $s(X^d)$ for some $d$ coprime with $M$.
With key-switching,
$\operatorname{RLWE}_{s(X^d)}\left(\mu(X^d)\right)$ is converted back to $\operatorname{RLWE}_{s(X)}\left(\mu(X^d)\right)$.

\subsection{Fast Computation of the Trace Operator}\label{sec:fast_trace}
The \emph{trace operator} is a fundamental concept in algebraic fields and structures that allows for the mapping of elements from a field extension back to the base
field. Trace operators play a crucial role in the analysis and manipulation of algebraic structures effectively, and they need to be computed efficiently to be used in optimization of packing algorithms (\cf \Cref{sec:server_ops}).
The trace operator of a polynomial $P(X)$ is given by $
\operatorname{Tr}\left(P(X)\right) = \sum_{\substack{1 \leq d \leq M \\ \gcd(d, M) = 1}} P\left(X^d\right) \mod \Phi_M(X)$. This can be homomorphically evaluated through $N$ key-switches. This is not only slow but also induces a lot of noise. To address this issue, one can use the factorization of the trace through partial traces. Recall that polynomials in $\mathcal{K} = \mathbb{Q}[X] \mod \mathbb{Z}[X] \mod \Phi_M$ form a Galois field extension of $\mathbb{Q}$ and, that there exists a towering field extension $\mathbb{Q} = \mathcal{K}_0 \subset \mathcal{K}_1 \subset \ldots \subset \mathcal{K}_{\alpha} = \mathcal{K}$. Therefore, denoting $\mathcal{K}_{i+1}$ as the Galois field extension of $\mathcal{K}_{i}$, and letting $\operatorname{Tr}_{\mathcal{K}_{i}/\mathcal{K}_{j}}$ be the partial trace from $\mathcal{K}_i$ down to $\mathcal{K}_j$ for $i>j$, and using the fact that $\operatorname{Tr}_{\mathcal{K}_1/\mathcal{K}_0}$ can be further factorized, \citet{cryptoeprint:2025/488} conceived an algorithm that homomorphically computes the complete trace with $(\alpha-1) (t -1) + \sum_{\ell\in \omega_{t-1}} (\ell-1)$ key-switches, where $M$ is as defined in \Cref{sec:enc_dec}, and $\omega_{s}$ is the set of all prime factors of any natural number $s$ counted with their multiplicity.

\subsection{Packing}\label{sec:packing}
The \emph{packing} operation combines several \ac{LWE} ciphertexts encrypting messages $\mu_i \in T_p$ into a single \ac{RLWE} ciphertext that encrypts a polynomial whose coefficients are the $\mu_i$s.
We use \ac{RLWE} to reduce communications overhead when exchanging ciphertexts.
This technique was first introduced in the setting of power-of-two cyclotomic polynomials and later extended to prime-power cyclotomic fields~\citep{cryptoeprint:2025/488}.
\emph{Fast packing} algorithms transform the input \ac{LWE} ciphertexts into an \ac{RLWE} ciphertext such that the first coefficient of the encrypted polynomial corresponds to the coefficient of interest. However, in this process, all other coefficients are randomized. The homomorphic trace operator is then applied to zero out all but the relevant coefficient. The resulting ciphertexts are subsequently rotated and summed to obtain a single ciphertext encrypting all the $\mu_i$s.  
By exploiting the decomposition property over towering Galois expansions, \citet{cryptoeprint:2025/488} showed how partial trace operations can be shared among different ciphertexts. This optimization reduces the number of required key-switching operations to $\sum_{\ell \in \omega_{t-1}} (\ell-1) \;+\; \sum_{1 \leq i \leq \alpha} \frac{1}{t^{i-1}}$ per packed coefficient, where $t$, $\alpha$, and $\omega_{t-1}$ are as defined in Sec.~\ref{sec:enc_dec} and \ref{sec:fast_trace}.

\section{System and Threat Model}\label{sec:sys_and_threat_model}
We now describe our system and threat model, and list the assumptions made in this work.

\textbf{\ac{ML} Model.}
The server stores an \ac{ML} model parameterized with weights $\theta$.
Each model weight is stored in plaintext and only known by the server.
Since we are using a \ac{TFHE} scheme 
which supports only integer arithmetic, we consider the case where the model weights and the activations (or outputs) are quantized. 
Crucially, the accumulators, which store intermediate sums of multiply–accumulate (MAC) operations also operate under predefined integer bit-widths.
To achieve this, the server could apply state-of-the-art \ac{QAT} techniques that enforce fixed-point integer representations for the weights, the activations and the accumulator~\citep{colbert2024aq,colbert2023a2q,wrapnet}. 
We refer to the bit-width of the accumulator as $b$ (\eg, 12-bit precision).
The bit-widths of weights and activations are typically lower than the accumulator (\eg 4-bit precision)~\citep{colbert2024aq}.
All the bit precisions are public information and known to clients.
Training a quantized model with 4-bit weights, 4-bit activations and 14-bit accumulator can maintain accuracy within 1\% of the original floating-point model~\citep{colbert2024aq}.

\textbf{Clients.}
The client has some input data \eg an image that it wants to evaluate using the server's model $\theta$.
The client generates a unique session identifier for each inference request.
All server-client messages carry this session identifier and a monotonic round counter; messages from stale or mismatched rounds are rejected.
We assume that the client remains online during the inference request in order to receive and process the intermediate outputs by the server, and to send back the processed outputs to the server.
Thus, \sys requires increased client participation across multiple short communication rounds compared to traditional client-server \ac{FHE} where the client only needs to be available to send the encrypted input data and collect the inference results from the server.
In practice, we find that each round typically completes within seconds, making transient disconnections easily mitigated by simple session resumption.
Furthermore, this assumption is consistent with prior secure inference frameworks~\citep{gazelle,delphi,shechi}.

\textbf{Threat Model.}
Our security objective is two-fold.
On the one hand, we want to preserve \emph{data privacy}, preventing the server from inferring information from the client input data that is used for inference.
On the other hand, we want to preserve \emph{model confidentiality}, preventing the client from learning about the model weights $\theta$.
Model confidentiality is important in client-server deployments because the model often represents proprietary intellectual property and investment due to high development and training costs.
Exposing model weights to clients could allow them to replicate or reverse-engineer the service, undermining monetization and intellectual property protections.

We assume that the client and server follow the \ac{FHE} protocol. However, we assume that the server is honest-but-curious, \ie it faithfully runs the protocol yet attempts to learn about the client inputs from all messages it sees.
On the other hand, we assume that the clients can be adversarial \ie they try to reverse engineer $\theta$.
To achieve this, they may send arbitrary input to the server in every round.
Network traffic is authenticated and encrypted (\eg, using TLS) and we do not consider side-channel leakage outside the protocol, such as timing or power measurements on client hardware.
Our explicit leakage is limited to model metadata disclosed during the setup phase (\cf \Cref{sec:protocol_setup}) and the size and number of messages sent between the server and client.

\section{Design of \sys}
\label{sec:protocol}
We now describe the design of \sys.
We first present the high-level workflow in \Cref{sec:nutshell} and then explain each step in detail in the remaining sections.

\begin{figure*}[t!]
\centering
\includegraphics[width=\textwidth]{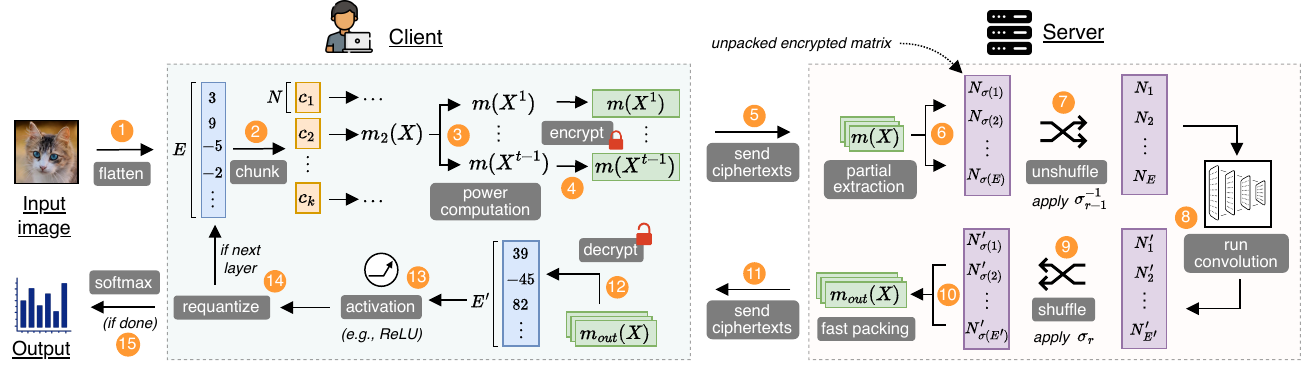}
\caption{The workflow of the \sys during an inference request using a \ac{CNN} model. The box on the left and right shows the operations performed on the client and server, respectively.
}
\label{fig:workflow}
\end{figure*}

\subsection{\sys in a Nutshell}
\label{sec:nutshell}

The core idea behind \sys is to split inference across the client and server: the server performs all linear operations (\eg convolutions, fully connected layers) under \ac{RLWE} encryption, while the client applies the nonlinearities (\eg ReLU) in plaintext. By letting the client periodically decrypt intermediate results, our hybrid design prevents the noise accumulation that would otherwise require costly bootstrapping on the server. Moreover, evaluating nonlinearities in the clear avoids the overhead of polynomial approximations used in conventional \ac{FHE}-based inference. Together, these choices yield substantial speedups over prior schemes.  
This design, however, introduces a new challenge: intermediate plaintext outputs become visible to the client, creating the risk of reverse-engineering the model weights. To protect model confidentiality, \sys applies a secret, random permutation to the server’s outputs before sending them back to the client. This obfuscation makes it practically infeasible for the client to reconstruct model parameters.

\Cref{fig:workflow} illustrates the \ac{E2E} workflow for an inference request in \sys. 
The inference proceeds for $ L $ rounds, corresponding to each model layer.
At the start of round $ r $, the client has a flattened representation of the input image (when $ r = 1 $, step 1) or the output of the previous layer (when $ r > 1 $).
The client encrypts these polynomials into multiple \ac{RLWE} ciphertexts and sends them to the server (steps 2-5).
The server then prepares the input matrix for the convolution by unpacking the coefficients in the \ac{RLWE} ciphertexts using an optimized \emph{partial extraction} scheme (step 6). 
Before running the linear operations, the server unshuffles ciphertexts to undo a secret permutation from the previous round, which is required for model confidentiality (step 7). 
The server then runs the convolution, shuffles the columns of the convolution output, and executes a fast packing algorithm to convert this output into multiple \ac{RLWE} ciphertexts (steps 8-10).
These ciphertexts are sent back to the client (step 11).
The client decrypts these ciphertexts, applies the activation in clear and requantizes the outputs for the next layer (steps 12-14), and repeats. 
Finally, when $r = L$, the client optionally runs softmax in clear to obtain the final output (step 15).

\subsection{Protocol Setup}
\label{sec:protocol_setup}

The protocol starts by the client generating an \ac{RLWE} encryption key $ s(X) $ and a key-switching key $\operatorname{KSK}$.
The key-switching key for automorphism $X \rightarrow X^d$ is given by $\operatorname{RLWE}_{s(X)}\left( \frac{s(X^d)}{D^i} \right)$ for $1\le i \le l$, where $l$ and $D$, representing \emph{depth of decomposition} and \emph{base}, respectively, are the parameters widely used in the context of homomorphic modular products~\citep{cryptoeprint:2025/488}.
The client computes the key-switching key for at most $\alpha(t-1)$ indexes.
The client then requests various elements it requires to process the client-side operations, including the required input shape for the first layer, all other layers input and output size, the specifications of the activation functions used (\eg, ReLU or sigmoid), and the per layer re-quantization scale $\eta$.
This exchange only has to be done once for each unique model $ \theta $ at the server and its communication overhead is negligible.

\subsection{Client Operations (pre-server)}
\label{sec:client_ops_design}

Here, we outline the operations performed by the client in a given round before communicating with the server.

\textbf{Flatten and chunk.}
For an input image $ I \in \mathbb{Z}^{H \times W \times C_{\operatorname{in}}} $, where $H$, $W$, and $C_{\operatorname{in}}$ represent the kernel height, kernel width, and number of input channels, respectively, is first flattened into a fixed order (\eg, row‑major with channels last) to obtain a one‑dimensional vector $ V \in \mathbb{Z}^E$
(step 1 in \Cref{fig:workflow}).
The client then splits $ V $ into $ k $ chunks $ c_0,\ldots, c_{k-1} $ each of length $ N $, and applies zero-padding to the final chunk $ c_{k-1} $ (step 2).
Each chunk $c_j=(c_{j,0},\ldots,c_{j,N-1})$ is hence embedded as the coefficients of a polynomial $m_j(X)$, \ie $m_j(X) = \sum_{i=0}^{N-1}c_{j,i}X^i $.

\textbf{Power computations.}
Next, the client computes, for each polynomial  $m(X)$, evaluations at specific powers of $X$ (step~3). The range of these powers is determined by the \emph{trace extraction level} parameter $\gamma$. After receiving the encrypted powers from the client, the server performs \emph{fast packing} (\cf \Cref{sec:server_ops}) at trace level $\gamma $, thereby significantly reducing the number of key switches per packed element. For example, if \( \gamma = 0 \), the client does not compute any powers and simply sends an encrypted version of each $m(X)$ to the server. If $ \gamma = 1 $, the client additionally computes and sends $ m(X^j) $ for all $ j \in \{1,\ldots,t-1\} $. For $ \gamma = 2 $, the client computes $ \nu(X)=m(X^{j(kt+1)}) $ for all $ j \in \{1,\ldots,t-1\}, k\in \{0,\ldots,t-1\} $. Each $ \nu(X) $ is then encrypted using a private $s(X) $ as $ \operatorname{RLWE}_{s(X)}(\nu(X)) $ (step~4) and sent to the server. 

\subsection{Server Operations}\label{sec:server_ops}
We next describe the operations performed by the server in a given round $r$.

\textbf{Partial trace extraction.}
If $\gamma>0$, for each encrypted polynomial chunk $\operatorname{RLWE}_{s(X)}\left(m_j(X)\right) $, using its associated encrypted powers computed by the client,
the server extracts an \ac{RLWE} encryption of $\operatorname{Tr}_{\mathcal{K}_\gamma /\mathcal{K}_0} \left( (m_j(X)\overline{\Omega}_i^*(X)\right)$ for $ 0 \le i \le N-1$, where $ \mathrm{Tr}_{\mathcal{K}_0 /\mathcal{K}_0}$ is the identity mapping and $\overline{P}(X) = P\left(X^{M-1}\right)$ for any $P \in \mathcal{K}$  (step 6).
It is worth noting that, as
$c_{j,i} = \langle m_j(X), \Omega_i^*(X) \rangle = \mathrm{Tr}\left( m_j(X) \overline{\Omega}_i^*(X)\right)$, where $c_{j,i}$ is as defined in \Cref{sec:client_ops_design},
each of the extracted \ac{RLWE} can be mapped to an encryption of one of the input coefficients using the homomorphic evaluation of the partial trace $\mathrm{Tr}_{\mathcal{K} / \mathcal{K}_{\gamma}}$. For a given $\gamma$, the partial trace is extracted using the expressions derived in the following lemma. In the interest of space, the proof is postponed to Appendix~\ref{app:lemma_proof}.

\begin{lemma}\label{lem:partial_trace_extraction}
For all $P \in \mathcal{K}$ and $ i=0,\ldots N-1$, we have:

$$
\mathrm{Tr}_{\mathcal{K}_\gamma /\mathcal{K}_0} \left( (P(X)\overline{\Omega}_i^*(X)\right)\nonumber =\begin{cases}
    \sum_{\substack{k=1}}^{\substack{t-1}}  P(X^k)\overline{\Omega}_i^*(X^k) \text{ for $\gamma=1$}\nonumber\\
    \sum_{\substack{j=0}}^{\substack{t-1}} \sum_{\substack{k=1}}^{\substack{t-1}}P(X^{k (jt+1)})\overline{\Omega}_i^*(X^{k (jt+1)})\text{ for $\gamma=2$}
\end{cases}
$$

\end{lemma}

As addition is homomorphic and the powers of the message sent by the client are encrypted, the only part that needs to be addressed now is the multiplication by the clear polynomial.
Recalling the expression of dual basis, one can derive an expression for $\overline{\Omega}_i^*(X^d)$. 
However, these polynomials do not have integer coefficients, and multiplication by a clear-text polynomial $P(X)$ is homomorphic if and only if $P(X) \in \mathbb{Z}[X]$. To alleviate this problem, the server instead multiplies by $M^{-1} \left( M\overline{\Omega}_i^*(X^d) \right)$ where $M^{-1} M = 1 \mod p$, where $p$ is as defined in \Cref{sec:preliminaries}.
Moreover, as these polynomials have only two non-zero coefficients, the multiplication can be cheaply performed by rotating and summing the clear-text, taking $4M$ elementary operations for one \ac{RLWE} encryption. 

\textbf{Unshuffle.}
Before the server runs each encrypted linear layer, it applies an unshuffle operation $\sigma_{r-1}^{-1}$ to the incoming unpacked encrypted matrices $ N_{\sigma(1)}, ..., N_{\sigma({E})} $ to undo the secret permutation $\sigma_{r-1}$ that was applied to the previous layer’s outputs when they were packed and returned to the client (step 7).
As we will discuss in \Cref{sec:model_privacy}, this shuffling and unshuffling is necessary for model confidentiality. 
For the first linear layer ($r=1$) there is no preceding permutation, so we set $\sigma_0^{-1} = \mathrm{id}$; consequently, the first unshuffle is a no‑op.
This results in the rows $ N_{1}, ..., N_{E} $ of the unpacked, unshuffled encrypted matrix $ N $ with shape $ E \times 2M$.

\textbf{Convolution.}
The server reshapes matrix $ N $ to obtain $ N_r $ with shape $ H\times W \times C_{\operatorname{in}} \times 2M $.
It then runs the convolution operation where $ 2M $ can be considered as batch size.
After reshaping, this results in matrix $ N' $ with dimension $ E'\times 2M $.

\textbf{Shuffle.}
Next, the server applies the permutation $\sigma_{r}$ to matrix $N'$, which shuffles its rows (step~9). 
The permutation $\sigma_{r}$ is generated uniformly at random from a secret seed derived from the session identifier and the current round number, ensuring it is unique for each round and client. 
The resulting matrix has rows $N'_{\sigma(1)}, \ldots, N'_{\sigma(E)}$.

\textbf{Fast Packing.}
The \emph{fast packing} operation takes as input $t^{\beta-1} (t-1)$ \ac{RLWE} ciphertext encryptions, for $1\leq \beta \leq \alpha -1$, of $ \operatorname{Tr}_{\mathcal{K}_\gamma / \mathcal{K}_0} \left(\overline{\Omega}_0^*(X) m'_{i} + B(X)\right) $ where $B \in \operatorname{ker} (\operatorname{Tr}_{\mathcal{K} / \mathcal{K}_\gamma})$
and outputs
$\operatorname{RLWE}_{s(X)}(m'(x))$, where $ m'(X) = \sum_{i=0}^{t^{\beta-1}(t-1)-1} m'_i X^{i \frac{M}{t^\beta}}$ (step 10).
Thus, the packing operation reduces the number of ciphertexts by a factor $t^{\beta-1} (t-1)$.  Finally, the resulting chunk-polynomials $m'(X)$ are sent to the client (step 11).

Packing is a computationally expensive operation, dominating runtime (\cf \Cref{fig:single_threaded_time_breakdown}) due to the key-switching operations needed for automorphisms. \Cref{tab:ksk_per_key} shows the number of key switches per packed element, which grows with the size of the prime $t$ and is lower bounded by $2.5$.
As a result, packing becomes a key factor in runtime and provides an important lever in selecting appropriate key sizes.
Applications must balance packing cost (favoring smaller $t$) against precision and security, as larger keys improve security but increase memory usage and slow basic operations.

\begin{table}
\centering
\footnotesize
\caption{\label{tab:ksk_per_key} Key switches per output coefficient for different start partial trace extraction level $(\gamma)$.}%
\begin{tabular}{lccc}
\toprule
\textbf{Key size} & \multicolumn{3}{c}{\textbf{Trace extraction level ($\gamma$)}}\\
$M = t^\alpha$ & $\gamma=0$ (Base) & $\gamma=1$ & $\gamma=2$ \\
\midrule
\textbf{$3^7$} & 2.50 & 1.50 & 0.50 \\
\textbf{$5^5$} & 3.25 & 1.25 & 0.25 \\
\textbf{$7^4$} & 4.16 & 1.16 & 0.16 \\
\bottomrule
\end{tabular}
\end{table}

However, thanks to the partial trace extraction operation, the ciphertexts that our fast packing algorithm receives as inputs in step 3 already encrypt the image of the desired polynomial by the partial trace $\mathrm{Tr}_{\mathcal{K}_\gamma/\mathcal{K}_0}$. This means that the fast packing procedure does not have to homomorphically compute the trace $\mathrm{Tr}_{\mathcal{K}_\gamma/\mathcal{K}_0}$, thus resulting in noticeable computational gains.  
From \Cref{tab:ksk_per_key}, we see that setting $\gamma=1$, \ie skipping the $\mathrm{Tr}_{\mathcal{K}_1/\mathcal{K_0}}$ operation,  makes the number of key switches per packed element bounded between $1$ and $1.5$ for any choice of $t$, thus relaxing the aforementioned trade-off.
Setting $\gamma = 2$, \ie skipping both $\mathrm{Tr}_{\mathcal{K}_1/\mathcal{K}_0}$ and $\mathrm{Tr}_{\mathcal{K}_2/\mathcal{K}_1}$ allows to further improve this result. 
At the cost of a bit more computation from the client and with more data transfers, the server can significantly reduce the cost of packing. We have investigated this trade-off in our experiments (\cf \Cref{sec:exp_single_thread}).

\subsection{Client Operations (post-server)}
Upon receiving the packed ciphertexts $\operatorname{RLWE}_{s(X)}(m’(X))$, the client decrypts each chunk to recover a polynomial $m'(X) $ whose coefficients contain a block of output entries.
Reading the designated packed positions and concatenating across all returned ciphertexts (and dropping any padding) yields the layer output vector $V' \in \mathbb{Z}^{E’} $.
The client then applies the layer’s activation function $ f $ in plaintext element-wise (\eg ReLU).
However, even when the weights and the inputs are quantized to low-bit values (\eg 4-bit), the output derived from several multiply-accumulate operations of a convolution can exceed this bit precision.
Therefore, the client requantizes the output to the predefined input precision for next layer: using the per-layer scale $\eta$, we map $ y \mapsto \mathrm{clip}\!\big(\lfloor y/\eta \rceil\big)$.
If this is the final layer of a classifier, the client directly computes the softmax over $V'$ to obtain probabilities and returns them to the user (step 15).
Otherwise, $V'$ becomes the input to the next round $r+1$ (step 14).

\subsection{Trade-offs}
At the core of \sys lies a deliberate architectural trade-off: we avoid costly bootstrapping altogether by performing intermediate decryptions on the client and executing non-linear operations in plaintext.
This significantly reduces the computational burden on the server, eliminating one of the main performance bottlenecks in \ac{FHE} inference.
However, this comes at the cost of additional network communication, as intermediate ciphertexts must be transmitted back and forth between the client and server after each linear block.
Thus, \sys exchanges the high compute latency of bootstrapping for higher communication volume, a trade-off that is particularly advantageous when network bandwidth is abundant but compute resources are at a premium.

\begin{table}[t]
\centering
\scriptsize
\setlength{\tabcolsep}{3pt}
\renewcommand{\arraystretch}{1.25}
\begin{threeparttable}
\caption{Design-level comparison of \textsc{Safhire} with existing approaches to private inference. $L$ denotes the number of linear blocks bounded by non-linearities.}
\label{tab:design-comparison}
\begin{tabularx}{\textwidth}{@{}>{\raggedright\arraybackslash}p{1.8cm}
  >{\raggedright\arraybackslash}X
  >{\raggedright\arraybackslash}X
  >{\raggedright\arraybackslash}X
  >{\raggedright\arraybackslash}p{1.3cm}
  >{\raggedright\arraybackslash}p{0.95cm}
  >{\raggedright\arraybackslash}p{1.05cm}
  >{\raggedright\arraybackslash}X
  >{\raggedright\arraybackslash}p{1.4cm}@{}}
\toprule
\textbf{Approach} &
\textbf{\ac{FHE} scheme} &
\textbf{Linear layers} &
\textbf{Non-linear layers} &
\textbf{Boot\-strapping required} &
\textbf{Client-server rounds} &
\textbf{Offline phase} &
\textbf{What the client learns} &
\textbf{Client threat model} \\
\midrule
Full encryption &
CKKS / BFV &
Plaintext--ciphertext products with slot rotations &
Polynomial approximations, server-side &
Yes &
1 &
No &
Final output only &
Malicious$^{a}$ \\
\addlinespace
Full encryption (TFHE) &
TFHE &
Scalar MACs with plaintext weights &
Exact, via programmable bootstrapping (lookup tables) &
Yes &
1 &
No &
Final output only &
Malicious$^{a}$ \\
\addlinespace
HE + 2PC &
HE + authenticated secret sharing &
HE on authenticated shares &
Garbled circuits with consistency checks &
No &
$O(L)$ &
Yes &
Final output only &
Malicious$^{a}$ \\
\addlinespace
Client-aided \ac{FHE} &
CKKS / TFHE &
Server-side under \ac{FHE} &
Client-side, in plaintext, exact &
No &
$O(L)$ &
No &
Intermediate pre-activations in the clear$^{b}$ &
Not addressed$^{b}$ \\
\midrule
\textbf{\textsc{Safhire} (ours)} &
RLWE &
Scalar MACs with plaintext weights &
Client-side, in plaintext, exact &
No &
$L$ &
No$^{c}$ &
Shuffled pre-activations (see \Cref{sec:model_privacy}) &
Malicious (crafted inputs) \\
\bottomrule
\end{tabularx}
\begin{tablenotes}[flushleft]\scriptsize
\item[$a$] The client only receives the final output; query-based model extraction from final outputs applies to every row and is considered beyond scope by all approaches.
\item[$b$] Without further protection, revealing intermediate pre-activations enables the exact weight recovery attack of Sec.~\ref{sec:model_privacy} (case (i)).
\item[$c$] Requires a one-time setup phase per model, involving key-switching keys and exchanging model metadata (see \Cref{sec:protocol_setup}).
\end{tablenotes}
\end{threeparttable}
\end{table} 
\subsection{Comparison with Existing Designs}
\Cref{tab:design-comparison} compares \sys with four families of private inference approaches.
\emph{Full encryption} approaches execute the entire forward pass on the server under \ac{FHE}.
CKKS- and BFV-based systems such as \textsc{CryptoNets}~\citep{cryptonets}, \textsc{LoLa}~\citep{brutzkus2019low}, \textsc{CHET}~\citep{chet}, \textsc{EVA}~\citep{eva}, \hyphen~\citep{hyphen}, and \orion~\citep{orion} pack values into SIMD slots, which requires slot rotations to evaluate convolutions, and replace activations with polynomial approximations, so deeper networks require bootstrapping.
\ac{TFHE}-based approaches~\citep{chillotti2021programmable,concreteml-resnet} instead evaluate activations exactly through programmable bootstrapping, at the cost of at least one bootstrapping operation per activation.
Both variants are non-interactive, requiring only a single round of client-server interaction, and the client only learns the final output.
\emph{HE + 2PC} protocols such as \textsc{Muse}~\citep{lehmkuhl2021muse} and \textsc{SIMC}~\citep{chandran2022simc} evaluate linear layers with HE and non-linear layers with garbled circuits with consistency checks.
This makes them secure against malicious clients, but requires an offline phase and multiple communication rounds per non-linear layer.
\emph{Client-aided FHE}, as supported by \textsc{nGraph-HE2}~\citep{boemer2019ngraph} and the hybrid mode of Concrete-ML~\citep{zama-hybrid}, lets the client evaluate non-linearities in plaintext.
This removes the need for bootstrapping, but exposes intermediate pre-activations to the client and, without further protection, enables exact recovery of the model weights (see \Cref{sec:model_privacy}).
\sys belongs to this last family, but differs in two ways.
First, its linear layers reduce to scalar MACs with plaintext weights on extracted ciphertexts, avoiding both slot rotations and bootstrapping.
Second, the server shuffles its outputs, so the client only observes permuted pre-activations, which protects the model against clients that send crafted inputs.
The price for these benefits is $L$ rounds of client-server interaction, compared to a single round for full encryption, and we quantify this trade-off in \Cref{sec:evaluation}.

\section{Model Confidentiality}
\label{sec:model_privacy}

We now analyze the model confidentiality guarantees of \sys. As outlined in \Cref{sec:sys_and_threat_model}, we assume that the model held by the server must remain private from the clients (model confidentiality).
Clients may send arbitrary inputs and use all server responses to infer the model weights.

To contextualize our contribution, we first establish a concrete attack vector. We present a novel method sufficient to demonstrate the system's practical vulnerabilities, thereby highlighting the crucial role server-side shuffling plays in $\sys$ to uphold model confidentiality. The key outcome of our presented attack is that it grants the adversary only partial information: the set of columns or rows of the target matrix, but not their canonical ordering. Consequently, the adversary faces the challenging reconstruction task of assembling the matrix from its components, which are permuted by an unknown function $\pi$. Crucially, to the best of our knowledge, there are no existing model-reconstruction attacks that function solely by exploiting a random ordering of the model weights. Furthermore, a brute-force search for the correct permutation is computationally infeasible given the size of our target matrix, \eg 512 columns. Therefore, \emph{even if an adversary successfully executes this attack, they cannot efficiently recover the original matrix.}

We start by showing how, in the absence of shuffling, the participating clients can deliberately craft and send inputs to the server that could compromise the confidentiality of the server-side model. Consequently, we illustrate how \sys remains robust against such adversarial model-inference attempts due to its integrated shuffling mechanism. We highlight that, to the best of our knowledge, we mount the most effective attack on \sys where the client aims to recover the server-side model.

\textbf{Attack Formulation.} Recalling that \sys focuses on linear server-side computations, fully connected and convolutional blocks can be written as $f(x)=Ax+b$ with $A\in\mathbb{R}^{m\times d}$ and $b\in\mathbb{R}^m$. A convolution is a matrix–vector product with the corresponding Toeplitz matrix~\citep{gray2006toeplitz}.
We compare three interface variants, and write $\sigma_{\operatorname{in}}$ and $\sigma_{\operatorname{out}}$ for unknown input and output permutations, which may be freshly sampled per query.
\emph{\textbf{(i) No shuffling}.} A client querying $x=\mathbf{0}$ reveals $b$. For each basis vector $e_i$, the response is $f(e_i)=Ae_i+b=a_i+b$, where $a_i$ is the $i$-th column of $A$; subtracting $b$ gives $a_i$. Repeating over every $i$, the client exactly recovers $A$ and $b$, thus compromising model confidentiality.
\emph{\textbf{(ii) Output shuffling only} (first round of \sys).} The server returns $\sigma_{\operatorname{out}}(Ax+b)$. The probe $x=\mathbf{0}$ reveals $b$ up to a permutation $\sigma_{\operatorname{out}}$ unknown to the client. To recover a column $a_i$ up to the same permutation, the client queries $x_1=e_i$ and $x_2=2e_i$, obtaining $y_1=\sigma'_{\operatorname{out}}(b+a_i)$ and $y_2=\sigma''_{\operatorname{out}}(b+2a_i)$. For any entry $\tilde{b}_k$ in the shuffled bias, there exists an index $j$ such that $(\tilde{b}_k+y_{2,j})/2$ appears in $y_1$, where $y_{2,j}$ is the $j$-th element of the vector $y_2$. Then $(y_{2,j}-\tilde{b}_k)/2$ equals the corresponding entry of $a_i$ at index $\sigma_{\operatorname{out}}^{-1}(k)$. Scanning all $k\in [m]$ reconstructs the multiset of entries of $a_i$ and iterating over $i\in [m]$ recovers all columns, but only \emph{up to a common unknown permutation}. If ties are present, they can be disambiguated with extra probes such as $3e_i$. Thus, even though output shuffling may reveal the values of the model weights, their ordering would remain private from the clients without the knowledge of the permutation function $\sigma_{\operatorname{out}}$. Therefore, in order to infer the exact indices of the model weights, the clients need to know $\sigma_{\operatorname{out}}$. To the best of our knowledge, no existing model-inversion-based attack in the literature can work just by exploiting an arbitrary ordering of the values of the model weights. 
In practice with $m=512$ in \resneteighteen, even with a brute force approach, the likelihood of guessing $\sigma_{\operatorname{out}}$ is $1/m!\approx 0$, thus making it practically impossible for the client to retrieve the exact model held by the server. 
\emph{\textbf{(iii) Input and output shuffling} (subsequent rounds of \sys).} The server computes $\sigma_{\operatorname{out}}(A\sigma_{\operatorname{in}} x + b)$, \ie an input $x=e_i$ is mapped to $e_{\sigma(i)}$ for some $\sigma(i)\in[m]$. Thus, the client cannot target a fixed column across queries without knowing
$\sigma_{\operatorname{in}}$ and $\sigma_{\operatorname{out}}$. Client-side observations have the form $\sigma_{\operatorname{out}}\!\left(b+\sum_{i} x_{\sigma_{\operatorname{in}}(i)} a_{\sigma_{\operatorname{in}}(i)}\right)$, which reveal only unordered mixtures; the individual columns $a_i$ cannot be isolated even up to permutation, and only the histogram of the weights of $A$ is revealed. Hence, the model-inversion-based reconstruction attacks are not applicable once input shuffling is applied. 

It is important here that the server does not shuffle the outputs from the last layer, as the client is expected to perform the \texttt{arg max} operation on the logits it receives from the server to conclude the classification-based inference. However, not shuffling the last layer's output still preserves the privacy of the server's model up to the unshuffling function, analogous to the input permutation in our analysis, of the preceding round. In other words, by observing the server's outputs in the last round, the client can learn the individual rows of $A$ but not their order up to the aforementioned input permutation of the previous round.

In addition to making it practically impossible for a client to exploit intermediate layer information to reconstruct the model stored on the server, we extend the advantage of shuffling the convolution outputs in every round to derive formal guarantees of \emph{\ac{DP}}~\citep{dwork2014algorithmic}. To this, we explore the \emph{shuffle model} of \ac{DP}~\citep{bittau2017prochlo,cheu2019distributed} and  
use the privacy amplification results of shuffling. For an input space $\mathcal{X}$ and output space $\mathcal{Y}$, the shuffle model of \ac{DP} involves a \emph{local randomizer} $\mathcal{R}:\mathcal{X}\mapsto \mathcal{Y}$ and a \emph{shuffler} $\psi: \mathcal{Y}^n\mapsto \mathcal{Y}^n$ for some $n\in\mathbb{N}$. $\mathcal{R}$ is responsible for obfuscating any $x\in \mathcal{X}$ by mapping it to some $y\in \mathcal{Y}$. For any set of messages $x_1,\ldots,x_n$ in $\mathcal{X}$ that has been point-wise obfuscated by $\mathcal{R}$, $\psi$ applies a random permutation $\sigma$ to the locally obfuscated messages $\mathcal{R}(x_1),\ldots,\mathcal{R}(x_n)$ to obtain $\mathcal{R}(x_{\sigma(1)}),\ldots,\mathcal{R}(x_{\sigma(n)})$. This essentially ensures that, for any $i=1,\ldots,n$, an observer cannot link a certain message $\mathcal{R}(x_{\sigma(i)})$ to its corresponding sender $i$ without knowing the permutation function $\sigma$ used by the shuffler. 

Recent studies~\citep{feldman2022hiding, erlingsson2019amplification, feldman2023stronger, biswas2023tight, koskela2023numerical, balle2019privacy} on shuffle model of \ac{DP} have shown that if $\mathcal{R}$ satisfies \emph{local} \ac{DP}, then the shuffling results in an amplification of the local \ac{DP} guarantees and $\psi \circ \mathcal{R}^{n}$ satisfies central \ac{DP} from the perspective of the observer with the corresponding amplified privacy bound. Aligned with this, in the following theorem, we formally derive the amplified \ac{DP} guarantees of the server's model under \sys. In the interest of space, the proof has been postponed to Appendix~\ref{app:DP_thm_proof}.

\begin{theorem}\label{th:DP_amp_gen}
    In each round of \sys, if the convolution outputs satisfy $(\varepsilon_0,\delta_0)$-local \ac{DP}, then for any $\delta\in[0,1]$ s.t. $\varepsilon_0\leq \ln\left(\frac{n}{16 \ln(2/\delta)}\right)$ the convolution outputs as a whole in each round, as observed by the client, satisfies $\left(\varepsilon,\delta+(e^{\varepsilon}+1)(e^{-\varepsilon_0}/2 +1)n\delta_0\right)$-\ac{DP}, where 
    $$\varepsilon\leq \ln\left(1+\frac{e^{\varepsilon_0}-1}{e^{\varepsilon_0}+1}\left(\frac{8\sqrt{e^{\varepsilon_0}\ln(4/\delta)}}{\sqrt{n}}+\frac{8e^{\varepsilon_0}}{n}\right)\right).$$
\end{theorem}

\begin{remark}\label{rem:DP_amplification}
We empirically observe that the noise introduced by the fast packing operation into each output of the convolutional layers follows a Gaussian distribution (see discussion below and \Cref{fig:error_histogram_plot}). Therefore, we empirically verify that the outputs of the convolution sent to the client by the server satisfy $(\varepsilon_0,\delta_0)$-local \ac{DP}. This, in turn, makes \Cref{th:DP_amp_gen} applicable to \sys and, hence, the outputs received by the client in any given round satisfy $(\varepsilon,\delta)$-\ac{DP} with an amplification in the privacy level as given by \Cref{th:DP_amp_gen}. 
\end{remark}

\textbf{Applicability of \Cref{th:DP_amp_gen}.} As outlined in \Cref{rem:DP_amplification}, \Cref{fig:error_histogram_plot} in Appendix~\ref{app:fhe_noise} provides an illustrative example showing that the noise induced by fast packing on the first layer outputs of \resnettwenty applied to \cifar indeed follows a Gaussian distribution.
We expect this behavior to carry over to subsequent layers as the fast-packing noise depends only on the packing parameters, which are identical across layers and models, whereas the noise contributed by the convolution itself is
small in comparison. These are empirical observations, not an analytical characterization; in particular, they do not bound the tails of the noise.
We reiterate that \Cref{th:DP_amp_gen} does not provide a theoretical proof that the convolution outputs in each round intrinsically satisfy local \ac{DP}. Formally instantiating a neighboring relation on the final model weights and showing that the fast-packing noise constitutes a calibrated local randomizer for it would additionally require group privacy over the affected coordinates due to the adjacency on the final model parameters, as well as composition across rounds and repeated queries, whose analyses remain orthogonal to the guarantees derived by \Cref{th:DP_amp_gen}. To the best of our knowledge, a rigorous theoretical characterization of the noise introduced by the fast packing operation remains an open problem and is beyond the scope of this paper. This is precisely why \Cref{th:DP_amp_gen} is stated conditionally, \ie the privacy amplification holds if the convolution outputs already satisfy local \ac{DP}, and \Cref{fig:error_histogram_plot} supports this precondition empirically rather than analytically.

\textbf{Injecting additional noise.}
    A noteworthy observation from \Cref{fig:error_histogram_plot} is that the noise added to the outputs of the convolutional layers is centered around $0$, and lies within the range $[-1/2p,1/2p]$. This ensures correct decryption on the client side, in accordance with the TFHE decryption procedure detailed in \Cref{sec:enc_dec}. Thus, the \ac{DP} amplification bound derived in \Cref{th:DP_amp_gen} incurs no additional loss in decryption correctness.  However, depending on the application-specific privacy requirements, the precondition of \Cref{th:DP_amp_gen} can be enforced by design. Before returning the outputs of a round, the server can add to each output coordinate fresh noise drawn from its own randomness, \eg discrete Gaussian noise on the message space $\mathbb{T}_p$~\citep{canonne2020discrete}, which can be added homomorphically as a plaintext without any key material. If this noise is calibrated so that each coordinate is $(\varepsilon_0, \delta_0)$-locally DP, then \Cref{th:DP_amp_gen} applies to the values decrypted by the client, within the scope stated in \Cref{rem:DP_amplification} and without relying on the empirical behavior of the fast-packing noise. This would, however, mean that the injected noise must exceed the decryption margin with non-negligible probability, so the client evaluates the non-linearities on perturbed pre-activations. This \ac{DP} guarantee by injecting additional noise would, therefore, come at a price of client-side decryption utility loss. \sys thus admits two operating points: exact inference, in which client-side decryption is exact and model confidentiality rests on shuffling, and noisy inference, in which \Cref{th:DP_amp_gen} yields a formal \ac{DP} guarantee at a cost in utility. Quantifying this privacy-utility trade-off, \eg by clipping the outputs to a narrower range before adding noise, is left for future work.

In summary, this section shows that shuffling prevents clients from reconstructing the server-side model in \sys. Even with adversarially crafted inputs, a client learns intermediate values only up to unknown permutations, which, to the best of our knowledge, existing reconstruction attacks cannot exploit. Privacy amplification by shuffling further yields and amplifies formal \ac{DP} guarantee for the outputs revealed to the client, conditional on these outputs being locally \ac{DP}. In scenarios requiring a stronger formal privacy guarantees, this \ac{DP} amplification can be made stronger and enforced by the server by injecting additional noise into the outputs shared with clients, albeit at the potential cost of reduced utility.

\section{Evaluation}
\label{sec:evaluation}

We conduct an experimental evaluation of \sys and answer the following questions:
\begin{enumerate}[label=\emph{\arabic*})]
    \item How does single-threaded \sys compare in terms of runtime against \orion, a state-of-the-art \ac{FHE} framework (\Cref{sec:exp_us_vs_orion})?
    \item What is the effect of different trace extraction and precision levels on the latency of individual \sys operations (\Cref{sec:exp_single_thread})?
    \item How does the runtime of \sys evolve when increasing the number of utilized CPU threads for different trace extraction and precision levels (\Cref{sec:exp_multi_threaded})?
    \item What is the runtime of \sys on a GPU for different trace extraction and precision levels (\Cref{sec:exp_gpu})?
\end{enumerate}

\begin{table}[t!]
\footnotesize
\centering
\caption{Parameterization of \sys for different precision levels $ b $ and levels of partial trace extraction level $ \gamma $.}
\label{tab:params_fhe}
\begin{tabular}{lcc|cc|cc}
\toprule
\textbf{Precision ($b$)} & \textbf{Key size} & \textbf{RLWE Noise ($\Delta$)} & \multicolumn{2}{c|}{\textbf{$\gamma = 0$}} & \multicolumn{2}{c}{\textbf{$\gamma \in [1, 2]$}} \\
                   &    $(M = t^\alpha)$   &       & $D$ & $l$ & $D$ & $l$ \\
\cmidrule(lr){4-5} \cmidrule(lr){6-7}
8 bits   & $3^7$           & $2^{-36}$      & 512        & 3          & 512        & 3         \\
12 bits  & $\mathbf{7^4}$  & $2^{-51}$      & $2^{12}$   & 3          & $2^{16}$   & 2         \\
16 bits  & $\mathbf{5^5}$  & $2^{-51}$      & 310        & 5          & 310        & 4         \\
\bottomrule
\end{tabular}
\end{table}

\subsection{Implementation}

We implement \sys, including its core cryptographic primitives, in the Julia programming language.
The core machine learning architecture relies on the \texttt{NNlib.jl} library and we leverage \texttt{FFTW.jl} for efficient Fast Fourier Transform computations during keyswitches.
We use the \texttt{Permutations.jl} library for handling permutation-based operations.
Our implementation supports multi-threaded and GPU execution.

To verify the correctness of our implementation, we conduct inferences across multiple precision levels $( b )$, partial trace extraction levels $( \gamma )$, and model–dataset pairs.
For each configuration, we generate random input vectors and perform inference using both a quantized cleartext model and \sys.
We then compare the resulting outputs element-wise.
Across all tested configurations, we observe no discrepancies between the cleartext and encrypted inference results, giving us high confidence in the correctness of our implementation.
This aligns with our empirical observations of the \ac{FHE} noise in Appendix~\ref{app:fhe_noise}.

\begin{table*}[t!]
\footnotesize
\centering
\begin{tabular}{|l|l|c||c|c|c|c|c|c|}
\hline
 \cellcolor{gray!20} & \cellcolor{gray!20} & \cellcolor{gray!20} & \multicolumn{6}{c|}{\cellcolor{gray!20}\textbf{Precision ($b$)}}\\
\cline{4-9}
\cellcolor{gray!20}\textbf{Model} & \cellcolor{gray!20}\textbf{\orion} &  \cellcolor{gray!20}\begin{tabular}
{@{}c@{}}\textbf{Extraction} \\ \textbf{level ($\gamma$)}\end{tabular} & \multicolumn{2}{c|}{\cellcolor{gray!20}\textbf{8 bit}} & \multicolumn{2}{c|}{\cellcolor{gray!20}\textbf{12 bit}} & \multicolumn{2}{c|}{\cellcolor{gray!20}\textbf{16 bit}}\\
\cline{4-9}
\cellcolor{gray!20} & \cellcolor{gray!20} & \cellcolor{gray!20} & \cellcolor{gray!30}\textbf{\sys} & \cellcolor{gray!20}\textbf{Speedup} & \cellcolor{gray!30}\textbf{\sys} & \cellcolor{gray!20}\textbf{Speedup} & \cellcolor{gray!30}\textbf{\sys} & \cellcolor{gray!20}\textbf{Speedup}\\

\hline \noalign{\vskip 0.1cm} \hline

\multirow{3}{*}{\begin{tabular}{@{}l@{}}\resnettwenty\\(\cifar)\end{tabular}} & \multirow{3}{*}{866.5 s} &  0 & 175.2 s & 4.9$\times$ & 277.1 s & 3.1$\times$ & 425.7 s & 2.0$\times$\\
 & & 1 & 135.6 s & 6.4$\times$ & 131.3 s & \textbf{6.6$\times$} & 247.4 s & 3.5$\times$\\
 & & 2 & 114.0 s & \textbf{7.6$\times$} & 252.8 s & 3.4$\times$ & 189.1 s & \textbf{4.6$\times$}\\

\hline \noalign{\vskip 0.1cm} \hline

\multirow{3}{*}{\begin{tabular}{@{}l@{}}\resneteighteen\\(\tinydataset)\end{tabular}} & \multirow{3}{*}{1253.9 s} & 0 & 633.1 s & 2.0$\times$ & 953.1 s & 1.3$\times$ & 1310.1 s & 1.0$\times$ \\
 & & 1 & 518.1 s & 2.4$\times$ & 521.2 s & \textbf{2.4$\times$} & 745.8 s & \textbf{1.7$\times$}\\
 & & 2 & 452.0 s & \textbf{2.8$\times$} & 776.5 s & 1.6$\times$ & 739.3 s & \textbf{1.7$\times$}\\

\hline \noalign{\vskip 0.1cm} \hline

\multirow{3}{*}{\begin{tabular}{@{}l@{}}\resnetthirtyfour\\(\imagenet)\end{tabular}} & \multirow{3}{*}{10819.6 s} & 0 & 4335.5 s & 2.5$\times$ & 6509.6 s & 1.7$\times$ & 9005.0 s & 1.2$\times$\\
 & & 1 & 3447.3 s & 3.1$\times$ & 3738.6 s & \textbf{2.9$\times$} & 5433.1 s & \textbf{2.0$\times$}\\
 & & 2 & 3164.0 s & \textbf{3.4$\times$} & 5544.3 s & 2.0$\times$ & 5298.0 s & \textbf{2.0$\times$}\\
 
\hline
\end{tabular}
\caption{The \ac{E2E} latencies of \sys over \orion and associated speedups, for different trace extraction levels and precisions. To compute the duration of network communication in \sys, we assume a conservative network speed of \SI{1.25}{\mega\byte\per\second}. For each model and precision, we mark the largest speedup in bold. 
}
\label{tab:orion_us_comparison}
\end{table*}

\subsection{Experimental Setup}
\label{subsec:exp_setup}

\textbf{Models and Datasets.}
We evaluate three representative \ac{CNN} models commonly used in prior \ac{FHE} studies~\citep{orion,lee2022privacy}: \resnettwenty (0.27M parameters), \resneteighteen (11.2M), and \resnetthirtyfour (21.8M)~\citep{he2016deep}.
We use three vision datasets, namely \cifar~\citep{krizhevsky2009learning} (with image size $32 \times 32$), \tinydataset~\citep{le2015tiny} (with image size $64 \times 64$), and \imagenet~\citep{deng2009imagenet} (with image size $224 \times 224$).
Images across all datasets have three input channels.
For efficiency evaluation in \sys, the semantic content of the images is immaterial; however, the input shape and the number of output classes impact inference time, as they determine the sizes of the first convolutional and final classification layers.
We use the standard ReLU function as activation function for \sys.
For \resnetthirtyfour on the \imagenet dataset, we replace the max pooling with average pooling, as the latter is linear and therefore significantly more efficient under \ac{FHE}. 

\textbf{\ac{FHE} Parameterization.}
\Cref{tab:params_fhe} presents the different parameters chosen in the \sys implementation, for different precision levels $ b $.
The noise, $D$, and $l$ are chosen to guarantee correct decryption of the ciphertext with high probability after each convolution and packing operation.
The key size is then chosen to compromise between size, which directly affect memory usage and runtime, and the number of key switches required for fast packing.
All parameter choices yield at least 128-bits of security when tested with M. Albrecht's security estimator~\citep{cryptoeprint:2015/046}.

\textbf{Baselines.}
We compare \sys against two baselines, one non-interactive and one interactive.
The non-interactive baseline is \orion~\citep{orion}, a state-of-the-art \ac{FHE} framework based on the CKKS scheme~\citep{cheon2017homomorphic}.
\orion evaluates the entire forward pass on the server in a single client-server round and optimizes inference by pairing single-shot multiplex packing with automated bootstrap placement.
However, unlike our hybrid \ac{RLWE} design, \orion keeps all layers encrypted and therefore still requires costly bootstrapping, as shown in \Cref{fig:motivation_bootstrapping_time}.
The interactive baseline is \concreteml~\citep{concreteml-resnet}, a high-performance open-source \ac{FHE} library based on the TFHE scheme, which we run in its hybrid inference mode.
In this mode, the server evaluates the linear layers under encryption, while the client decrypts their outputs, evaluates the activations in plaintext, and re-encrypts the results for the next layer.
\concreteml thus follows the same client-aided approach as \sys and, like \sys, requires a client-server round for every encrypted layer, which makes it an \emph{interactive} baseline: any speedup of \sys over it cannot stem from interactivity alone.
For \concreteml, we quantize the weights and activations of the served models to 6-bit precision, as we observe a significant degradation in its performance at higher precision.
\concreteml automatically allocates the accumulator bit-width of each layer, so it differs across layers.
Across our experiments, this value ranges from 13 to 17 bits.

\textbf{Compute Platform.}
We conduct our experiments on three distinct hardware platforms to evaluate CPU and GPU performance separately.
CPU-only evaluations are performed on two servers. The first is equipped with two AMD EPYC 7K62 CPUs with 2.60GHz base frequency, providing 192 logical cores total; 2003GB of RAM; and runs Ubuntu 20.04.
Due to the hardware requirements of \orion, a more powerful server was needed for experiments involving the \resnetthirtyfour model. This second machine, used exclusively for these tests, uses two Intel$^{\circledR}$ Xeon$^{\circledR}$ Platinum 8272L CPUs with a 2.60GHz base frequency and 104 logical cores; 2975GB of RAM; and also runs Ubuntu 20.04.
GPU-accelerated experiments are run on an on-demand compute cluster. Each experiment utilizes a node with 24 cores of an AMD EPYC 7543 CPU with a 2.80GHz base frequency and a single NVIDIA A100-SXM4 GPU with 80GB of VRAM. These nodes run Ubuntu 22.04.

\textbf{Metrics.}
Our evaluation focuses on two metrics:
\begin{enumerate*}[label=(\arabic*)]
    \item the wall-clock time for client- and server-side operations, and
    \item the total communication volume between the client and server.
\end{enumerate*}
The \ac{E2E} latency of \sys consists of computations on both the client and the server, and the ciphertext transfer time.
Network latency is highly dependent on the conditions in the deployment setting and we separately measure the overhead of communication under different network conditions.
Client-side operations include encryption, decryption, activations, and requantization, while server-side operations consist of packing, linear operations, and extraction.
As the client-side operations are orders of magnitude faster, we mainly focus on optimizing the overhead of server-side operations.
We report their individual times, as well as the total sum, which we refer to as \emph{server-side execution time}. We also report the accuracy achieved by \sys under different precision levels.

\subsection{Performance of \sys Compared to \orion}
\label{sec:exp_us_vs_orion}

\begin{table}[t!]
    \footnotesize
    \centering
    \begin{tabular}{|c|c|c|c|c|c|}
        \hline
       \cellcolor{gray!20}\textbf{Dataset}  & \cellcolor{gray!20}\textbf{Model} & \cellcolor{gray!20}\textbf{\concreteml} & \cellcolor{gray!20}\textbf{\orion (SiLU)} & \multicolumn{2}{c|}{\cellcolor{gray!30}\textbf{\sys}}  \\
       \cline{5-6}
        \cellcolor{gray!20} & \cellcolor{gray!20}& \cellcolor{gray!20}& \cellcolor{gray!20}& \cellcolor{gray!20}\textbf{16 bit} & \cellcolor{gray!20}\textbf{12 bit} \\
         \hline \noalign{\vskip 0.1cm} \hline
      \cifar & \resnettwenty & 91.46\% & 91.70\% & 90.00\% & 89.63\% \\
      \tinydataset & \resneteighteen & 52.37\% & 57.00\% & 53.19\% & 53.16\% \\
      \imagenet & \resnetthirtyfour & 64.06\% & N/A & 70.16\% & 49.47\% \\
      \hline 
    \end{tabular}
    \caption{Top-1 accuracy of \sys across configurations. 
    }
    \label{tab:accuracy}
\end{table}

First, we compare \sys against \orion on the same hardware for single-threaded execution.
\Cref{tab:orion_us_comparison} compares the \ac{E2E} latency against \orion under a conservative network speed of 1.25 \si{\mega\byte/s} (10 \si{\mega\bit/s}), for different extraction levels $ \gamma $ and precisions $ b $.
In this setting \sys offers significant speedups ranging from $1.0\times$ for $\gamma = 0$ and $1.6\times$ for $\gamma > 0$ up to $7.6\times$.
Increasing $b$ generally increases the runtime, but not monotonically, since each combination of $b$ and $\gamma$ requires a different parameterization, as we discuss in \Cref{sec:exp_single_thread}.
For a fixed model and $ b $, speedups increase when increasing $\gamma$ in majority of cases.
We recall that $\gamma = 0$ represents the baseline which does not use our extraction optimization which is a vital contribution of this work.

\Cref{tab:accuracy} reports the accuracy of \sys, \orion, and \concreteml.
We use A2Q+~\citep{colbert2024aq} to train the quantized models of \sys under the specified accumulator precision.
On \cifar, the accuracy of \sys is within $2.1\%$ of \orion and $1.8\%$ of \concreteml.
On \tinydataset, it is within $3.9\%$ of \orion and exceeds that of \concreteml by up to $0.8\%$.
As we use the default hyperparameters of A2Q+, careful tuning can narrow these gaps further~\citep{colbert2024aq}.
Obtaining the accuracy of \orion is considerably more costly than for \sys.
Whereas the encrypted outputs of \sys match those of the quantized cleartext model, \orion builds on CKKS, an approximate scheme in which encryption noise, rescaling, and bootstrapping perturb intermediate values throughout the computation.
Its accuracy can thus only be measured faithfully by running encrypted inference over the entire evaluation set, which is prohibitively expensive for \resnetthirtyfour at about three hours per image (\Cref{tab:orion_us_comparison}).
We therefore do not report the accuracy of \orion on \resnetthirtyfour, consistent with the paper introducing \orion~\citep{orion}.
On this model, we compare \sys only against \concreteml: \sys reaches up to $70.16\%$ accuracy, compared to $64.06\%$ for \concreteml.

\begin{figure}[t!]
	\centering
	\includegraphics{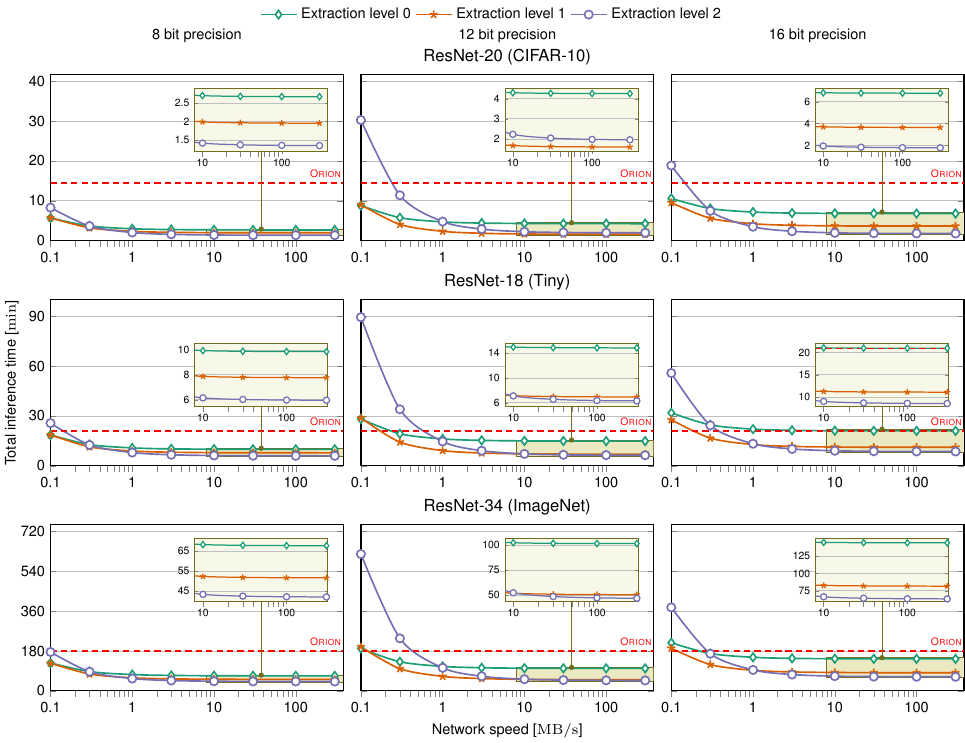}
	\caption{Total inference time for \sys under varying the network speed between the client and server. We consider networks speeds of 0.1, 0.3, 1, 3, 10, 30, 100, and 300 \si{\mega\byte/s}.}
	\label{fig:network_speed_time}
\end{figure}

We further explore the effect of network latency on the \ac{E2E} inference time in \Cref{fig:network_speed_time}.
The points on this plot are calculated by summing the computational times for the client and server with an approximated network transfer time.
This transfer time is derived by dividing the total exchanged data by a hypothetical bandwidth.
We select hypothetical bandwidths ranging from 0.1–300 MB/s such that they are roughly logarithmically spaced and representative of consumer connections.
For simplicity of presentation, we assume that the client and server have symmetrical upload and download speeds.

\Cref{fig:network_speed_time} shows that even for a network speed as low as \num{0.3} \si{\mega\byte/s}, the inference time of \sys is comparable or faster than \orion in most settings.
For all higher bandwidths, \sys with extraction optimization ($\gamma > 0$) outperforms \orion, even at modest network speeds.
These results also highlight that even at a network speed as low as \num{3} \si{\mega\byte/s}, the inference time for \sys is close to its performance at much higher speeds, emphasizing that computation, rather than communication, is the primary bottleneck beyond that point.
At high speeds, $\gamma=2$ yields little benefit over $\gamma=1$, while at low speeds, $\gamma=1$ clearly outperforms $\gamma=0$. Overall, $\gamma=1$ emerges as a robust choice across network regimes.

\begin{table}[t!]
    \centering
    \footnotesize
    \begin{tabular}{|c|c|c|c|c|c|}
        \hline
        \multicolumn{3}{|c|}{\cellcolor{gray!40}} & \multicolumn{3}{c|}{\cellcolor{gray!20}\textbf{Extraction level ($\gamma$)}}\\
        \cline{4-6}
        \multicolumn{3}{|c|}{\cellcolor{gray!40}} & \cellcolor{gray!20}\textbf{0} & \cellcolor{gray!20}\textbf{1} & \cellcolor{gray!20}\textbf{2}\\
        \hline
        \hline
        \cellcolor{gray!20} & \cellcolor{gray!20} & \cellcolor{gray!20}\textbf{Client} & 0.3 s & 0.4 s & 0.8 s \\
        \cellcolor{gray!20} & \cellcolor{gray!20}\multirow{-2}{*}{\textbf{8 bits}} & \cellcolor{gray!20}\textbf{Server} & 591.4 s & 465.6 s & 356.5 s \\
        \cline{2-6}
        \cellcolor{gray!20} & \cellcolor{gray!20} & \cellcolor{gray!20}\textbf{Client} & 0.3 s & 1.2 s & 5.7 s \\
        \cellcolor{gray!20} & \cellcolor{gray!20}\multirow{-2}{*}{\textbf{12 bits}} & \cellcolor{gray!20}\textbf{Server} & 888.8 s & 415.2 s & 371.6 s \\
        \cline{2-6}
        \cellcolor{gray!20} & \cellcolor{gray!20} & \cellcolor{gray!20}\textbf{Client} & 0.3 s & 1.8 s & 3.8 s\\
        \cellcolor{gray!20}\parbox[t]{2mm}{\multirow{-6}{*}{\rotatebox[origin=c]{90}{\textbf{Precision ($b$)}}}} & \cellcolor{gray!20}\multirow{-2}{*}{\textbf{16 bits}} & \cellcolor{gray!20}\textbf{Server} & 1257.5 s & 664.3 s & 508.6 s \\
        \hline
    \end{tabular}
    \caption{Breakdown of total inference time between client and server on \tinydataset with \resneteighteen.
    }
    \label{tab:client_server_op_time_breakdown}
\end{table}

\subsection{Single-threaded Performance of \sys}
\label{sec:exp_single_thread}

Next, we analyze how the different operations in \sys contribute to its performance for single-threaded execution for inference with a single sample.

\Cref{tab:client_server_op_time_breakdown} reports the split of total computational time between the client and the server for \tinydataset inference with the \resneteighteen model across different values of $ b $ and $ \gamma $. 
This table shows that the client-side execution time increases with the extraction level, yet it remains negligible compared to the server-side execution time, which is between $65\times$ and $3682\times$ greater across all tested settings. Therefore, for the rest of this section, we focus specifically on the server-side operations as they are the dominating factor in total inference latency.

\begin{figure}[t!]
	\centering
	\includegraphics{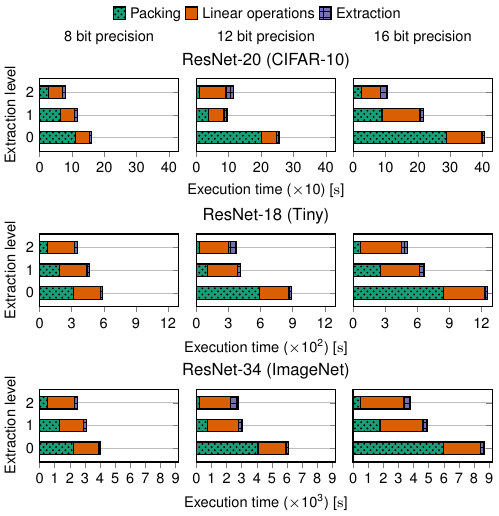}
	\caption{Breakdown of the server-side execution time for a single sample inference in \sys across various settings.}
	\label{fig:single_threaded_time_breakdown}
\end{figure}

\Cref{fig:single_threaded_time_breakdown} shows a time breakdown of server-side operations. 
We make two key observations.
First, increasing $ \gamma $ beyond 0 significantly reduces the time taken for packing, for all models and values of $ b $ but slightly increases the time required for extraction.
Second, linear operations require more time as $ b $ increases, particularly for \resneteighteen and \resnetthirtyfour, but remain constant as $ \gamma $ varies and $ b $ is fixed.

\begin{figure}[ht]
	\centering
	\includegraphics{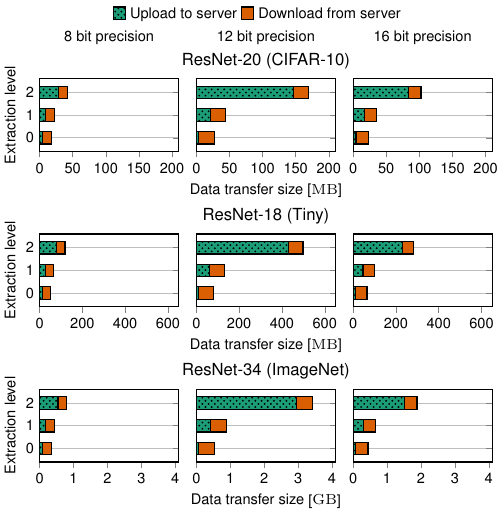}
	\caption{Breakdown of the client-server communication for a single sample inference in \sys across various settings.}
	\label{fig:single_threaded_data_breakdown}
\end{figure}

To further analyze the communication-computation trade-offs, we show the data transfer size to and from the server in \Cref{fig:single_threaded_data_breakdown}, for different models and values of $ b $ and $\gamma$.
For a fixed model and value of $ b $, the size of data sent by the server to the client remains constant across different values of $ \gamma $, but the size of uploaded data varies.
This is because for higher $ \gamma $, the client sends additional higher-degree polynomials to the server.
In exchange for this higher upload cost, increasing $ \gamma $ allows the server to spend significantly less time on packing and slightly more on extraction (see \Cref{fig:single_threaded_time_breakdown}).

Both \Cref{fig:single_threaded_time_breakdown} and \Cref{fig:single_threaded_data_breakdown} highlight the computation-communication trade-offs.
As discussed in \Cref{sec:client_ops_design}, this trade-off stems from the client providing the server with higher-degree polynomials for higher extraction level.
This results in fewer key switches for the server, as shown in \Cref{tab:ksk_per_key}, which significantly reduces packing time.
However, as can be seen, this also significantly increases the amount of data the client must upload.
For example, with 16-bit quantization on \cifar, increasing the extraction level from 0 to 2 reduces the server time spent on packing from \num{71}{\%} to \num{24}{\%} of the server-side execution time.
In the same setting, however, the data uploaded by the client increases from \num{4.2} to \num{83.9} \si{\mega\byte}.

Finally, we discuss why the runtime of \sys does not always grow with $b$.
Each combination of $b$ and $\gamma$ requires its own parameterization (\Cref{tab:params_fhe}): $b$ determines the key size $M = t^\alpha$ and, together with $\gamma$, the decomposition depth $l$.
These parameters affect the server-side operations differently.
The number of key switches per packed coefficient depends on $t$ and $\gamma$ (\Cref{tab:ksk_per_key}), and the cost of each key switch grows with $l$ and $M$.
Partial extraction evaluates $t-1$ terms per coefficient for $\gamma = 1$ and $t(t-1)$ terms for $\gamma = 2$ (\Cref{sec:client_ops_design}), while the cost of the linear operations depends on $M$ but not on $\gamma$.
For $\gamma = 0$, the cost of packing dominates and becomes more expensive with every increase in $b$. For example, $b = 12$ requires more key switches per packed coefficient than $b = 8$ (4.16 versus 2.50), and $b = 16$, while requiring fewer (3.25), uses a deeper decomposition ($l = 5$ versus 3) on a larger key.
For $\gamma = 1$, in contrast, $b = 12$ requires the fewest key switches (1.16 versus 1.50 for $b = 8$) with the shallowest decomposition ($l = 2$ versus 3), and thus outperforms $b = 8$ on all three models (\eg \SI{415.2}{\second} versus \SI{465.6}{\second} of server-side execution time on \resneteighteen, \Cref{tab:client_server_op_time_breakdown}).
For $\gamma = 2$, packing is cheap for every $b$, so the runtime is driven by the linear operations, which grow with $b$, and by partial extraction, which is most expensive for $b = 12$ (42 terms per coefficient versus 6 and 20 for $b = 8$ and $b = 16$).
For the two larger models, the linear operations dominate, and the runtime again increases monotonically with $b$, whereas on \cifar, $b = 12$ ends up slightly slower than $b = 16$ (\Cref{fig:single_threaded_time_breakdown}).

\subsection{Multi-threaded Performance of \sys}
\label{sec:exp_multi_threaded}

\begin{table*}[t!]
\centering
\footnotesize
\setlength{\tabcolsep}{2pt}
\begin{tabular}{|l|C{26.52pt}|C{26.52pt}|c||c|c|c|c|c|c|c|c|c|}
\hline
 \cellcolor{gray!20} & \multicolumn{2}{c|}{\cellcolor{gray!20}} & \cellcolor{gray!20} & \multicolumn{9}{c|}{\cellcolor{gray!20}\textbf{Precision ($b$)}}\\
\cline{5-13}
\cellcolor{gray!20}\textbf{Model} & \multicolumn{2}{c|}{\cellcolor{gray!20}\textbf{\concreteml}} &  \cellcolor{gray!20}\begin{tabular}{@{}c@{}}\textbf{Extraction} \\ \textbf{level ($\gamma$)}\end{tabular} & \multicolumn{3}{c|}{\cellcolor{gray!20}\textbf{8 bit}} & \multicolumn{3}{c|}{\cellcolor{gray!20}\textbf{12 bit}} & \multicolumn{3}{c|}{\cellcolor{gray!20}\textbf{16 bit}}\\
\cline{2-3} \cline{5-13}
\cellcolor{gray!20} & \cellcolor{gray!20}\begin{tabular}{@{}c@{}}\textbf{Time}\\(s)\end{tabular} & \cellcolor{gray!20}\begin{tabular}{@{}c@{}}\textbf{Data}\\(GB)\end{tabular} & \cellcolor{gray!20} & \cellcolor{gray!30}\begin{tabular}{@{}c@{}}\textbf{\sys}\\(s)\end{tabular} & \cellcolor{gray!20}\begin{tabular}{@{}c@{}}\textbf{Data}\\(MB)\end{tabular} & \cellcolor{gray!20}\textbf{Speedup} & \cellcolor{gray!30}\begin{tabular}{@{}c@{}}\textbf{\sys}\\(s)\end{tabular} & \cellcolor{gray!20}\begin{tabular}{@{}c@{}}\textbf{Data}\\(MB)\end{tabular} & \cellcolor{gray!20}\textbf{Speedup} & \cellcolor{gray!30}\begin{tabular}{@{}c@{}}\textbf{\sys}\\(s)\end{tabular} & \cellcolor{gray!20}\begin{tabular}{@{}c@{}}\textbf{Data}\\(MB)\end{tabular} & \cellcolor{gray!20}\textbf{Speedup}\\

\hline \noalign{\vskip 0.1cm} \hline

\multirow{3}{*}{\begin{tabular}{@{}l@{}}\resnettwenty\\(\cifar)\end{tabular}} & \multirow{3}{*}{86.8} & \multirow{3}{*}{1.8} & 0 & 60.9 & 18.0 & 1.4$\times$ & 109.7 & 27.2 & 0.8$\times$ & 147.4 & 22.7 & 0.6$\times$\\
 & & & 1 & 35.6 & 22.7 & 2.4$\times$ & 22.2 & 44.6 & 3.9$\times$ & 36.9 & 35.3 & 2.4$\times$\\
 & & & 2 & 19.7 & 41.7 & \textbf{4.4$\times$} & 16.0 & 169.9 & \textbf{5.4$\times$} & 20.5 & 102.5 & \textbf{4.2$\times$}\\

\hline \noalign{\vskip 0.1cm} \hline

\multirow{3}{*}{\begin{tabular}{@{}l@{}}\resneteighteen\\(\tinydataset)\end{tabular}} & \multirow{3}{*}{357.5} & \multirow{3}{*}{23.1} & 0 & 175.7 & 51.8 & 2.0$\times$ & 325.1 & 79.9 & 1.1$\times$ & 400.1 & 65.2 & 0.9$\times$\\
 & & & 1 & 84.6 & 65.1 & 4.2$\times$ & 63.6 & 131.0 & 5.6$\times$ & 109.7 & 99.7 & 3.3$\times$\\
 & & & 2 & 55.3 & 118.4 & \textbf{6.5$\times$} & 48.2 & 499.0 & \textbf{7.4$\times$} & 59.7 & 283.6 & \textbf{6.0$\times$}\\

\hline \noalign{\vskip 0.1cm} \hline

\multirow{3}{*}{\begin{tabular}{@{}l@{}}\resnetthirtyfour\\(\imagenet)\end{tabular}} & \multirow{3}{*}{2116.9} & \multirow{3}{*}{32.1} & 0 & 1181.0 & 349.3 & 1.8$\times$ & 2145.8 & 540.2 & 1.0$\times$ & 2840.3 & 436.2 & 0.7$\times$\\
 & & & 1 & 638.9 & 439.4 & 3.3$\times$ & 443.9 & 892.2 & 4.8$\times$ & 632.4 & 664.7 & 3.3$\times$\\
 & & & 2 & 337.4 & 800.1 & \textbf{6.3$\times$} & 307.7 & 3427.2 & \textbf{6.9$\times$} & 378.9 & 1883.1 & \textbf{5.6$\times$}\\
 
\hline
\end{tabular}
\caption{The computation latencies of \sys over \concreteml on multi-threaded execution and associated speedups, for different trace extraction levels and precisions. Both solutions are run using 32 threads. Latencies cover the client and server computation without network transfer, and Data is the total amount of data exchanged between the client and the server per inference. For each model and precision, we mark the largest speedup in bold.
}
\label{tab:concreteml_us_comparison}
\end{table*}

\begin{figure}[ht]
	\centering
	\includegraphics{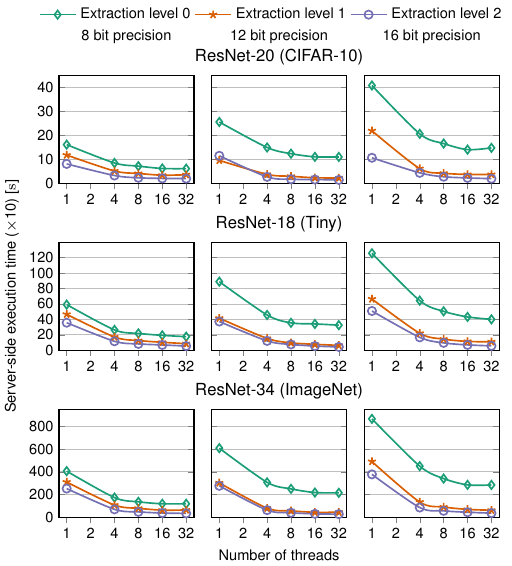}
	\caption{The breakdown of server-side execution time of \sys in function of the number of utilized threads.}
	\label{fig:data_transfer_latency_plot_tiny_multithreading_breakdown}
\end{figure}

Next, we quantify the benefits of multithreading on the server-side performance of \sys.
For these experiments, we measure the total server-side execution time while utilizing 1, 4, 8, 16 and 32 threads.
\Cref{fig:data_transfer_latency_plot_tiny_multithreading_breakdown} shows the results of these measurements across various precisions $b$ and extraction levels $\gamma$.
Notably, for $b=8$ server-side execution time stays close across different values of $\gamma$. Meanwhile, for $b=12$ and $b=16$ the gap between $\gamma = 0$ is significantly higher than for $\gamma=1$ and $\gamma=2$, which are relatively comparable to each other.

To put 32-thread results into perspective, we provide a direct comparison against hybrid model inference with \concreteml in \Cref{tab:concreteml_us_comparison}. We focus this comparison on 32-thread execution, as \concreteml is a natively multithreaded implementation and is neither designed nor optimized for single-threaded operation.
For both we report cummulative compoutation latency across the server client combined along with the data exchanged between the two in the process.
We observe that \sys with extraction optimization ($\gamma > 0$) achieves between $2.4\times$ and $7.4\times$ computational speedup over \concreteml in the same settings, while exchanging one to two orders of magnitude less data.

Moreover, multithreading provides substantial computational speedup over \orion: up to $54.0\times$ on \cifar, up to $26.0\times$ on \tinydataset, and up to $35.2\times$ on \imagenet. Higher extraction levels benefit from the highest relative speedup. For example with $b=8$, on \cifar, the multithreading speedup grows from $2.63\times$ ($\gamma=0$) to $4.10\times$ ($\gamma=2$), on \tinydataset $3.37\times$ ($\gamma=0$) to $6.45\times$ ($\gamma=2$), and finally on \imagenet $3.44\times$ ($\gamma=0$) to $7.48\times$ ($\gamma=2$). Precision plays a lesser role, though higher $b$ yields slightly larger gains (\eg, $4.09\times$ for $b=8$ vs. $5.93\times$ for $b=16$ on \cifar).

\begin{figure}[ht]
	\centering
	\includegraphics{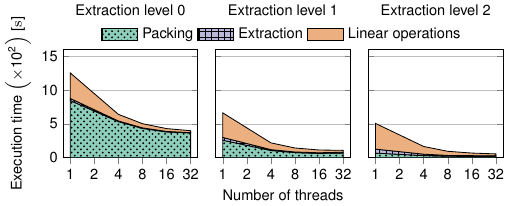}
	\caption{Breakdown of the duration of individual server operations when increasing the number of threads, for \resneteighteen with the \tinydataset dataset, while fixing $ b = 16 $.}
	\label{fig:multithreading_breakdown}
\end{figure}

Scaling up to 16 threads delivers the largest improvements; beyond that, benefits taper off ($<$5\% difference between 16 and 32 threads), with occasional slowdowns arising likely from the overheads of thread management.
Furthermore, \Cref{fig:multithreading_breakdown} breaks down how the execution time of individual operations scales with an increasing number of threads, using \resneteighteen with $b=16$ on the \tinydataset dataset. These measurements reveal that every operation benefits from multithreading. In this configuration, we observe up to $14.10\times$ speedup for linear operations and between $2.31\times$ and $4.09\times$ speedup for packing, compared to a single-threaded execution.
Overall, \sys benefits strongly from multithreading, especially at $\gamma=1$ and $\gamma=2$, while gains beyond 16 threads are marginal.

\subsection{Accelerating \sys with a GPU}
\label{sec:exp_gpu}

\begin{figure}[t!]
	\centering
	\includegraphics{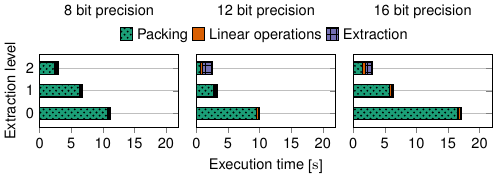}
	\caption{The breakdown of server-side execution time on A100 GPU for a single sample inference using a \resnettwenty model on \cifar dataset.
    }
	\label{fig:gpu_cifar10_resnet20}
\end{figure}

Finally, we enhance the \sys implementation and offload compute-intensive operations such as key switches, packing and partial extraction to the GPU.
\Cref{fig:gpu_cifar10_resnet20} breaks down the execution time for different extraction levels $ \gamma $, quantization levels $ b $ and when using the \cifar dataset with a \resnettwenty model, when running \sys on an A100 GPU.
We remark that the communication volume overhead is similar to the one shown in \Cref{fig:single_threaded_data_breakdown} (top).
For $ \gamma = 0 $ and $ b = 8 $, a single inference request takes \SI{11.3}{\second} of compute time using a GPU which is a speedup of 15.5$\times$ and compared to a single-threaded CPU setting which takes \SI{175.2}{\second}.
In line with other experiments, we observe that the execution time reduces as $ \gamma $ and $ b $ increase, at the cost of additional communication volume.
For all configurations of $ \gamma $ and $b $, we notice that the packing operation requires the most server-side compute time, \eg, for $ \gamma = 0 $ and $b = 16$ the packing operation takes 96.1\% of all compute time.
Additional GPU-specific optimizations, such as custom kernels, could further reduce this cost.
In summary, \sys benefits substantially from GPU acceleration, making GPUs a promising path for practical FHE inference.

\section{Related Work}
\label{sec:related_work}

\ac{FHE} based neural-network inference has progressed from early proofs of concept to GPU-accelerated systems at ImageNet scale. We group prior work into six strands and position our RLWE-based hybrid within this landscape.

\textbf{Leveled \ac{FHE} and Compiler Tool-Chains.}
\textsc{CryptoNets} demonstrated end-to-end neural network inference on encrypted MNIST using leveled homomorphic encryption, replacing non-linearities with the square function as the activation~\citep{cryptonets}. 
However, \textsc{CryptoNets} incurred high latency. 
Follow-up work addressed both accuracy and efficiency: \textsc{CryptoDL} achieved higher accuracy by retraining networks with carefully chosen low-degree polynomial approximations of common activations~\citep{cryptodl}, while \textsc{LoLa} reduced latency substantially through optimized data layout and alternating ciphertext representations~\citep{brutzkus2019low}. Compiler stacks such as \textsc{CHET} and \textsc{EVA} further automated parameter selection, layout, and (where applicable) bootstrapping for CKKS, improving developer productivity while matching or exceeding hand-tuned baselines~\citep{chet,eva}. These systems remain fully encrypted and therefore pay for polynomial activations and/or bootstrapping.

\textbf{Interactive and Hybrid Cryptographic Protocols.}
To reduce online latency, several works combine HE with secure two-party computation. \textsc{Gazelle} evaluates linear layers under CKKS and non-linearities via garbled circuits, achieving sub-second online times on small CNNs but requiring multiple interactive rounds and a semi-honest two-party model~\citep{gazelle}. \textsc{Delphi} refines this MPC/HE split by front-loading rotation/packing costs in a preprocessing phase to further shrink the online path~\citep{delphi}. More recently, \textsc{Shechi} introduced the first multiparty homomorphic encryption (MHE) compiler, automatically translating Python code into secure distributed computation that combines homomorphic encryption with multiparty computation~\citep{shechi}. 
\textsc{Shechi} focuses on distributed analytics such as PCA and genomic workloads, showing up to $15\times$ runtime improvements over prior secure frameworks and highlighting the promise of compiler-based optimization for hybrid cryptographic protocols.
Our approach also splits work across cryptographic boundaries, but is \emph{non-interactive}: non-linearities execute locally on the client, avoiding online MPC while still protecting server-side parameters.

\textbf{Scalable Fully Encrypted Frameworks.}
Recent frameworks push fully encrypted inference to deeper networks. \textsc{HyPHEN} introduces GPU-friendly kernels (RAConv/CAConv) and weight-reuse techniques to bring single-GPU ResNet-18/ImageNet latency to the tens-of-seconds regime~\citep{hyphen}. \textsc{Orion} adds single-shot multiplex packing and automated bootstrap placement, outperforming earlier CKKS baselines on ResNet-20 and demonstrating the first \ac{FHE}-based YOLO inference under CKKS~\citep{orion}. \textsc{EncryptedLLM} shows that small GPT-style models can run end-to-end under \ac{FHE} with large GPU/CPU speedups~\citep{encryptedllm}. All these systems remain fully homomorphic and thus still pay for bootstrapping or high-degree polynomial activations.

\textbf{Client–Server Hybrid Execution.}
Zama's \concreteml library exposes hybrid models as a developer feature, allowing certain layers to run on the client in plaintext and the rest under encryption on the server~\citep{concreteml-resnet}.
However, it remains at the level of developer tooling rather than a systematically studied design point. 
In contrast, our work formalizes the hybrid setting on an RLWE backend, introduces partial-trace packing and other optimizations to reduce key-switch pressure, and provides a systematic evaluation against state-of-the-art \ac{FHE} framework on diverse neural models and datasets. 
Moreover, our work explicitly analyzes and mitigates model leakage via randomized shuffling with differential-privacy amplification, dimensions which are unexplored in existing hybrid toolkits.

\textbf{Quantization and Low-Precision Accumulators.}
Quantization is orthogonal yet complementary to HE. \textsc{WrapNet} adds overflow-penalty regularization and cyclic activations to maintain accuracy with very low-precision arithmetic~\citep{wrapnet}. \textsc{A2Q}/\textsc{A2Q+} constrain weight norms during training to avoid accumulator overflow, recovering near-baseline accuracy with 14–16-bit accumulators on ImageNet-scale networks~\citep{colbert2023a2q,colbert2024aq}. We leverage these insights by selecting 8/12/16-bit accumulator widths compatible with our discrete-torus plaintext space.

\textbf{Bootstrapping Advances in \ac{TFHE}.}
Programmable bootstrapping over torus has seen large constant-factor improvements and enabled early CNN experiments under gate-by-gate evaluation~\citep{tfhe2018,tfhefastboot}. 
Our design circumvents bootstrapping entirely, leveraging client-side computation and trading extra network round-trip for substantial speed-ups.

\section{Conclusion}
\label{sec:conclusion}

We presented \sys, a practical hybrid \ac{FHE} inference framework that eliminates bootstrapping by offloading non-linear operations to the client while keeping linear layers encrypted on the server.
This design drastically reduces computation cost, avoids approximation of non-linearities, and ensures model confidentiality through shuffling.
Our evaluation on standard \acp{CNN} shows that \sys significantly lowers E2E latency compared to \orion, a state-of-the-art \ac{FHE} inference system.
Overall, \sys demonstrates that our hybrid approach can make privacy-preserving \ac{ML} inference both efficient and practical.

\section*{Acknowledgements}

This work has been funded by the Swiss National Science Foundation, under the project ``FRIDAY: Frugal, Privacy-Aware and Practical Decentralized Learning'', SNSF proposal No. 10.001.796.
The authors thank Mathis Randl for constructive discussions, and the anonymous reviewers and the action editor of TMLR for their valuable feedback and time. 

\bibliography{references/main}
\bibliographystyle{tmlr}

\appendix

\section{Putting It Together: One Round of \sys}\label{app:fhe-intuition}

We summarize one round of \textsc{Safhire} (Figure~\ref{fig:workflow}) in terms of the primitives above.
\begin{enumerate}
  \item \emph{Client.} The client flattens the layer input and splits it into polynomials $m_j(X)$ of $N$ coefficients (steps~1-2). It computes $m_j(X^d)$ for the $[\mathcal{K}_\gamma : \mathcal{K}_0]$ automorphisms of $\mathrm{Tr}_{\mathcal{K}_\gamma/\mathcal{K}_0}$ (step~3), encrypts all these polynomials with \ac{RLWE} under $s(X)$, and sends them (steps~4-5).
  \item \emph{Server.} For each input coefficient $c_{j,i}$, the server computes an encryption of $\mathrm{Tr}_{\mathcal{K}_\gamma/\mathcal{K}_0}\big(m_j(X)\, \overline{\Omega}^*_i(X)\big)$ using only additions and multiplications by integers and integer polynomials (step~6, \Cref{lem:partial_trace_extraction}), so that no key switch is needed. By applying the remaining levels $\mathrm{Tr}_{\mathcal{K}/\mathcal{K}_\gamma}$ to its plaintext would yield $c_{j,i}$. The server then undoes the permutation of the previous round (step~7), evaluates the linear layer as integer multiply--accumulate operations on these ciphertexts (step~8), and permutes the outputs (step~9). Finally, it completes the trace with automorphisms and key switching and packs the outputs into ciphertexts of elements of $\mathcal{K}_\beta$ (step~10), which it returns (step~11).
  \item \emph{Client.} The client decrypts by rounding, reads the outputs at the exponents $i\, t^{\alpha-\beta}$ (step~12), applies the activation, and requantizes (steps~13--14). The noise accumulated during the round is removed by the rounding step of decryption, and the next round starts from fresh encryptions.
\end{enumerate}
\section{Discussion}
\label{sec:app:discussion}

\paragraph{Computational discrepancy between the client and the server.}
In MLaaS deployments, the server typically has abundant computational resources, whereas client hardware is heterogeneous and outside the provider's control. By offloading non-linearities to the client, \sys departs from the fully server-side paradigm and requires the client to remain online throughout an inference request (\Cref{sec:sys_and_threat_model}), which proceeds in $L$ rounds (\Cref{sec:nutshell}). This is an explicit trade-off: a client that is slow to respond or temporarily disconnects increases the E2E latency of its request, whereas in fully encrypted inference the client only needs to be available to submit its input and retrieve the result.

At the same time, the work that \sys offloads to the client is precisely the work that is expensive under FHE but cheap in plaintext. Under full encryption, non-linearities require high-degree polynomial approximations or programmable bootstrapping, and ciphertext noise must be periodically reset through bootstrapping, which alone accounts for the majority of \orion's inference time (\Cref{fig:motivation_bootstrapping_time}). In \sys, the client instead evaluates the activations exactly in plaintext, and its intermediate decryption and re-encryption reset the ciphertext noise. The parameterized linear layers, which account for the bulk of the arithmetic and benefit the most from hardware acceleration (\Cref{sec:exp_multi_threaded,sec:exp_gpu}), are still evaluated under encryption on the server. This split is highly asymmetric: in a $3\times3$ convolution over 64 input channels, each output entry requires 576 multiply-accumulate operations, which the server performs on ciphertexts, but only a single activation, which the client evaluates in plaintext. Consequently, the client needs no accelerator. On a single CPU thread, the client-side computation for \resneteighteen on \tinydataset takes between 0.3\,s and 5.7\,s per inference, which is at least $65\times$ lower than the corresponding server-side execution time (\Cref{tab:client_server_op_time_breakdown}). Since this cost grows with the trace extraction level~$\gamma$, the choice of $\gamma$ also provides a way to accommodate weaker clients: with $\gamma = 0$, the client encrypts its input polynomials $m(X)$ without evaluating them at additional powers of~$X$ (\Cref{sec:client_ops_design}) and spends about 0.3\,s per inference, at the price of more packing work on the server (\Cref{fig:single_threaded_time_breakdown}).

Moreover, client unavailability delays a request without wasting the work already performed. For a bounded period of time, the server can retain the outputs of the last completed round, which remain encrypted under the client's key, together with the session identifier and round counter (\Cref{sec:sys_and_threat_model}). It re-sends them once the client reconnects, after which inference resumes with the next round. Finally, a session waiting for its client occupies memory but no compute, leaving the server to use the unutilized compute to process the rounds of other clients in the meantime. A slow or disconnected client therefore delays only its own request.

\paragraph{Relation to federated learning.}
Federated learning~\citep[FL;][]{federated_learning} is a distributed learning paradigm in which multiple clients collaboratively train a shared model under the coordination of a central server, without pooling their data. At first glance, its system model resembles ours: both involve a server and one or more clients that interact over multiple rounds to perform a machine learning task while aiming to keep sensitive data hidden from the other participants, server and clients alike. The two settings, however, serve fundamentally different purposes: FL is designed for training, whereas \sys targets inference. They also differ in what must be protected. In FL, the sensitive information is the clients' training data. It must remain confidential from both the server and the other clients, since both observe functions computed from it, such as model updates. The model itself is shared with all participants, and its weights change from round to round as training progresses. In \sys, by contrast, the sensitive information is twofold: the client's inputs, which must remain hidden from the server, and the model weights, which must remain hidden from the client. Moreover, the weights are fixed once the model is deployed. Finally, although the server may serve multiple clients concurrently, each client interacts with it in isolation, so no computation ever combines data from different clients. The inter-client leakage that is a fundamental concern within FL therefore does not apply to our setting.
\section{Table of Notation}
\label{sec:app:table_of_notation}

We list in \Cref{tab:params_names} the key notations used in this paper.

\begin{table}[h]
\centering
\begin{tabular}{lp{5.3cm}}
\toprule
\textbf{Param.} & \textbf{Description} \\
\midrule
$\mu$ & A cleartext message\\
$\operatorname{LWE}_{s}(\mu)$ & \ac{LWE} encryption of $\mu$ with key s\\
$\operatorname{RLWE}_{s(X)}(\mu(X))$ & \ac{RLWE} encryption of polynomial $\mu(X)$ with key s(X)\\
$M$ & Cyclotomic polynomial index; size of \ac{RLWE} key \\
$N$ & $\phi(M)$ s.t. $\phi$ is Euler's totient function\\
$\alpha$ & Exponent in $M = t^\alpha$ \\
$t$ & Prime base in $M = t^\alpha$ \\
$\Delta$ & Std. dev. of Gaussian encryption noise\\
$p$ & Size of cleartext message\\%
$D$ & Decomposition base\\
$l$ & Depth of the decomposition\\
$r$ & Round number in \sys\\
$\beta$ & Packing level \\
$\gamma$ & Partial trace extraction level\\
$b$ & The bit width of model accumulator \\
$\theta$ & Model held by the server\\
$I$ & Input image by the client\\
$H$ & Height of input image $ I $\\
$W$ & Width of input image $ I $\\
$ C_{in} $ & Channels of input image $ I $\\
$L$ & Number of layers in $\theta$\\
$\eta$ & Re-quantization scale\\

\bottomrule
\end{tabular}
\caption{\label{tab:params_names} Relevant RLWE parameters}
\end{table}
\section{Proof of \Cref{lem:partial_trace_extraction}}
\label{app:lemma_proof}
    Let $P \in \mathcal{K}$ and $0 \le i \le N-1$.
    Using Equations (23) and (24) from \citet{cryptoeprint:2025/488}, for all $Q\in \mathcal{K}$, we obtain:
    \begin{align}
        &\mathrm{Tr}_{\mathcal{K}_1 /\mathcal{K}_0}(Q(X)) = \sum_{\substack{i=1}}^{\substack{t-1}} Q(X^i)\label{eq:partial_trace_1_proof}\\
        &\mathrm{Tr}_{\mathcal{K}_2 /\mathcal{K}_1}(Q(X)) = \sum_{\substack{i=0}}^{\substack{t-1}} Q(X^{it+1})\label{eq:partial_trace_2_proof}
    \end{align}
    Substituting $Q(X)$ for $\overline{\Omega}_i^*(X)P(X)$ in \eqref{eq:partial_trace_1_proof} yields the first result, and
    composing this with \eqref{eq:partial_trace_2_proof} yields the second.%
    \qed

\section{Proof of \Cref{th:DP_amp_gen}}\label{app:DP_thm_proof}
    Let, in any given round, the convolution outputs satisfy $(\varepsilon_0,\delta_0)$-local DP. Then, by the very design of \sys, the convolution outputs of each layer are shuffled by the server, independently in each round, before being sent to the client. Therefore, the result follows from Theorem 3.8 of \citet{feldman2022hiding}.
\qed
\section{Implementation}
\label{sec:implementation}

\begin{figure}[t!]
	\centering
	\includegraphics{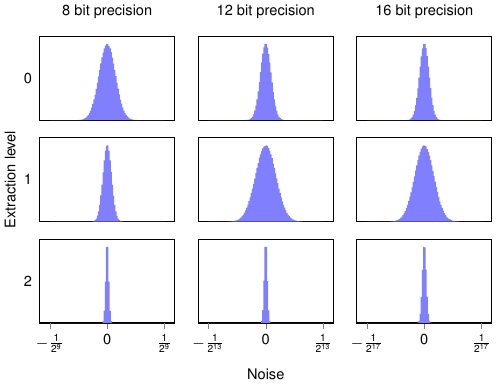}
	\caption{The distribution of noise from the fast packing operation for varying precision and extraction levels.}
	\label{fig:error_histogram_plot}
\end{figure}

\subsection{\ac{FHE} Noise in \sys}
\label{app:fhe_noise}

\Cref{fig:error_histogram_plot} shows the noise distribution induced by fast packing on the outputs of the first layer of \resnettwenty when applied to the \cifar dataset, for different extraction levels $\gamma$ and precision levels $ b $.
As long as the noise falls within the indicated boundaries, the decryption is correct.
For the given combinations of $\gamma $ and $ b $, we observe that the noise level remains within the allotted boundaries.
While we cannot formally exclude pathological cases, our empirical evidence strongly suggests that decryption failure does not occur in practice for our parameter choices.

\end{document}